\documentclass{ametsocV6.1}
\nolinenumbers

\title{A Generalized Richardson Number Diagnostic for Turbulence in the Free Atmosphere}

\authors{Mohamed Foudad,\aff{a}\correspondingauthor{Mohamed Foudad, m.foudad@reading.ac.uk} 
Miguel A.C. Teixeira,\aff{b} 
Paul D. Williams,\aff{a}
and Thorsten Kaluza\aff{c}  
}
\affiliation{\aff{a}{Department of Meteorology, University of Reading, Reading, United Kingdom}\\
\aff{b}{LAETA, IDMEC, Instituto Superior Tecnico, Universidade de Lisboa, Lisbon, Portugal}\\
\aff{c}{Deutscher Wetterdienst, Offenbach, Germany}
}

\abstract{A new Richardson number formulation, $Ri_{\mathrm{new}}$, is introduced to improve the diagnosis of turbulence in the stratified free atmosphere. The formulation is derived from the turbulent kinetic energy budget and  accounts for both vertical wind shear and horizontal shear (deformation and divergence), weighted by the ratio of horizontal to vertical eddy viscosities ($K_{mh}/K_{mv}$). This extends the classical Richardson number $Ri_{\mathrm{old}}$, which accounts only for vertical shear. The diagnostics $Ri_{\mathrm{new}}$, $Ri_{\mathrm{old}}$, and the widely used Turbulence Index~1 (TI1), computed from ERA5 reanalysis, are evaluated using more than 247 
million automated turbulence reports from commercial aircraft (2017--2024). Across various turbulence intensity thresholds, $Ri_{\mathrm{new}}$ consistently outperforms the other diagnostics, resulting in higher AUC values and improved probability of detection at operationally relevant false-alarm rates. The highest skill is obtained for $K_{mh}/K_{mv} \approx 5000$. Seasonal and regional evaluations indicate that the added 
value of $Ri_{\mathrm{new}}$ is largest where turbulence generation involves both vertical and horizontal shear, such as over the contiguous United States and during summer. $Ri_{\mathrm{new}}$ remains the best-performing diagnostic in all regions and seasons. Spatial case studies show that $Ri_{\mathrm{new}}$ identifies 83--98\% of observed moderate-or-greater turbulence events compared with 54--85\% for $Ri_{\mathrm{old}}$. This substantial improvement in detection comes with a much smaller increase in false alarms, confirming that $Ri_{\mathrm{new}}$ provides a more physically realistic representation of turbulence-prone regions. These results 
demonstrate that incorporating horizontal wind shear into 
the Richardson number yields a physically consistent and 
statistically robust improvement in turbulence diagnostics, with relevance for research and operational applications.}

\begin{document}

\maketitle

%
%
%
\statement

Turbulence at cruising altitudes is a major aviation hazard and is expected to increase as the climate warms. The Richardson number is a classical measure of the balance between turbulence suppression by atmospheric stability and generation by vertical wind shear. Here, we introduce an extension that also accounts for horizontal wind shear, which can be important near jet streams. Using more than 247 million aircraft turbulence reports, we show that this formulation consistently improves turbulence detection compared to the traditional Richardson number and a widely used turbulence index. The improvement is robust across turbulence intensities, seasons, and regions. These findings provide a more physically consistent approach to diagnosing turbulence and may support advances in aviation forecasting and climate impact assessments. 
%
%
%


\section{Introduction}

Turbulence in the free atmosphere remains one of the most significant challenges for both atmospheric research and aviation operations \citep{wyngaard2010,sharman2016nature}. Clear-air turbulence (CAT), in particular, occurs in cloud-free regions, making it difficult to identify visually or detect with conventional remote sensing instruments. CAT is primarily generated by instabilities within the stably stratified upper troposphere and lower stratosphere, often associated with strong vertical and horizontal wind shear in the vicinity of jet streams, upper-level fronts, and gravity wave breaking \citep[e.g.,][]{Dutton,Ellrod1992,knox1997,lane2004,sharman2006,sharman2012, kim2010, lee2018}. For aviation, turbulence represents a major operational and safety hazard, causing injuries, aircraft damage, and economic losses due to rerouting and flight delays \citep{sharman2016,gultepe2019,foudad2026}. This challenge is expected to intensify in the coming decades, as several studies project increases in both the frequency and intensity of clear-air turbulence under climate change, associated with enhanced upper-tropospheric wind shear \citep[e.g.,][]{williams2013, williams2017, kim2023, foudad2024}.

Despite decades of research, diagnosing and forecasting turbulence from large-scale atmospheric fields remains challenging. Turbulence occurs on spatial and temporal scales much smaller than those resolved by current numerical weather prediction (NWP) models. Direct forecasting of turbulence at these scales is therefore not feasible, now or in the foreseeable future \citep{sharman2017}. 
Instead, forecasting relies on diagnostics derived from resolved synoptic- and mesoscale atmospheric fields with grid spacings of tens of kilometers that serve as proxies for the likelihood of subgrid-scale turbulence. The multiscale nature of atmospheric turbulence has long been recognized. \cite{richardson1922} famously described the cascade of energy across scales, noting that ``big whirls have little whirls, which feed on their velocity, and little whirls have lesser whirls, and so on to viscosity.'' Following this, many diagnostics have been developed---ranging from empirical to physically based indices---to diagnose regions of high turbulence potential by assuming a downscale energy cascade from the resolved large scales to the aircraft scales \citep[e.g.,][]{brown1973,colson1965,dutton1980,Ellrod1992,kaplan2006,knox2008,mccann2012, bechtold2021, ko2025}.

Among the physically based diagnostics, the gradient Richardson number ($Ri$) is perhaps the most fundamental, as it characterizes the balance between stabilizing buoyancy and destabilizing shear production \citep{Stull1988}, and has become one of the most widely used diagnostics of turbulence in both research and operational forecasting \citep[e.g.,][]{endlich1964,Dutton}. The traditional $Ri$ can be interpreted as arising from a one-dimensional (1D) approximation of the turbulent kinetic energy (TKE) budget under the assumption of horizontal homogeneity. This leads to the classical formulation:

\begin{equation}
Ri = \frac{N^{2}}{S_{v}^{2}},
\end{equation}

where $N^2$ is the square of the Brunt-V\"{a}is\"{a}l\"{a} frequency, a measure of static stability, and $S_{v}^{2}$ is the square of the vertical shear of horizontal wind.

Values of $Ri$ below the theoretical critical threshold of 0.25 are associated with dynamically unstable conditions conducive to Kelvin--Helmholtz instability and turbulence generation \citep{miles1961,miles1964}. Although $Ri$ has shown some skill as an indicator of CAT occurrence \citep[e.g.,][]{mccann2001}, its performance remains limited in the free atmosphere \citep[e.g.,][]{dutton1980,Ellrod2015,lee2022,kaluza2025}. We believe that this limitation arises partly because the classical formulation considers only vertical wind shear and neglects horizontal velocity gradients, such as deformation and divergence, that also contribute to shear production in the full TKE budget. As a result, $Ri$ may underestimate the likelihood of turbulence generation in flows where three-dimensional shear plays an important role.

This limitation motivates the development of an extended Richardson number diagnostic that accounts for the full three-dimensional structure of the velocity field. In this study, we introduce a new Richardson number formulation, $Ri_{\mathrm{new}}$, derived directly from the full TKE budget equation, in which turbulent fluxes are parameterized using a first-order turbulence closure. This formulation explicitly accounts for both vertical and horizontal shear contributions to TKE production. The generalization incorporates the effects of horizontal deformation and divergence, weighted by the ratio of horizontal to vertical eddy viscosities, and therefore provides a more complete measure of the balance between stratification and three-dimensional shear production. The diagnostic skill of $Ri_{\mathrm{new}}$ is evaluated using ERA5 reanalysis and a large global dataset of in situ aircraft turbulence observations, and compared with the classical Richardson number and a commonly used turbulence index, which is considered one of the best-performing diagnostics of upper-level turbulence.

The remainder of this paper is organized as follows. Section~2 presents the derivation of $Ri_{\mathrm{new}}$ from the TKE budget. Section~3 describes the data and verification metrics. Section~4 evaluates the performance of $Ri_{\mathrm{new}}$, and Section~5 provides a summary and discussion.

\clearpage
\newpage

\section{Derivation of the new Richardson number}
\label{sec:theory}
\subsection{The TKE budget equation}

The turbulent kinetic energy (TKE) budget equation for a stratified flow is derived from the incompressible Navier-Stokes equations under the Boussinesq approximation \citep[e.g.,][]{Stull1988}. Following Reynolds decomposition, each flow variable is expressed as the sum of a mean and a turbulent fluctuation, e.g. $u_i = \overline{u_i} + u_i'$, where $i=1,2,3$ correspond to the zonal, meridional, and vertical directions respectively, and repeated indices imply summation. The TKE is defined as $\overline{e} = \frac{1}{2}\overline{u_i'u_i'}$, and its budget takes the form \citep{Stull1988}:

\begin{equation}
\frac{\partial \overline{e}}{\partial t}
+ \overline{u_j}\frac{\partial \overline{e}}{\partial x_j}
= \delta_{i3}\frac{g}{\overline{\theta}}\overline{u_i'\theta'}
- \overline{u_i'u_j'}\frac{\partial \overline{u_i}}{\partial x_j}
- \frac{\partial \overline{u_j'e}}{\partial x_j}
- \frac{1}{\overline{\rho}}\frac{\partial \overline{u_i'p'}}{\partial x_i}
- \varepsilon
\label{eq:tke}
\end{equation}

where $g$ is the gravitational acceleration, $\overline{\theta}$ is the mean potential temperature, $\delta_{i3}$ is the Kronecker delta, $\overline{\rho}$ is the mean density, and $\varepsilon$ is the viscous dissipation rate. The terms on the right-hand side represent, respectively, buoyancy production/destruction, shear production, turbulent transport, pressure correlation, and viscous dissipation of TKE \citep{Stull1988}.

The terms on the right-hand side of \eqref{eq:tke} that directly 
govern the generation or suppression of turbulence --- and that 
will be used in the derivation of $Ri_{\mathrm{new}}$ --- are the 
buoyancy term and the shear production term.

The buoyancy term $B$ is defined as
\begin{equation}
B = \frac{g}{\overline{\theta}}\,\overline{w'\theta'}
\label{eq:B}
\end{equation}
which is positive under unstable stratification 
($\overline{w'\theta'} > 0$, turbulence generation) and negative 
under stable stratification ($\overline{w'\theta'} < 0$, 
turbulence suppression), the latter being typical of the free 
atmosphere outside regions of deep convection.

The shear production term $SP$ is defined as
\begin{equation}
SP = -\overline{u_i'u_j'}\,\frac{\partial 
\overline{u_i}}{\partial x_j}
\label{eq:SP}
\end{equation}
which represents the transfer of kinetic energy from the mean flow to turbulent fluctuations. It is generally positive in the presence of wind shear and therefore acts as a source of turbulence.

\subsection{The new Richardson number $Ri_{\mathrm{new}}$}

A common approach to estimate whether turbulence will be generated or suppressed in a stratified flow is to define a dimensionless ratio between the stabilizing and destabilizing mechanisms, known as the Richardson number ($Ri$). This ratio reflects the balance of the TKE budget: stable stratification is associated with a decrease in TKE, whereas instability leads to increased TKE. If the destabilizing effects exceed the stabilizing effects, turbulence is maintained; otherwise, turbulence decays \citep{Stull1988, wyngaard2010}.

In a stably stratified atmosphere, the buoyancy term $B$ tends to suppress turbulence (stabilizing factor), while the shear production term $SP$ tends to generate turbulence (destabilizing factor). The ratio of these two terms defines the flux Richardson number $Ri_f$:

\begin{equation}
Ri_f = -\frac{B}{SP} = 
\frac{\dfrac{g}{\overline{\theta}}\,\overline{w'\theta'}}
{-\overline{u_i'u_j'}\,\dfrac{\partial \overline{u_i}}
{\partial x_j}}
\label{eq:Rif}
\end{equation}

Here we do not assume horizontal homogeneity, contrary to 
standard practice, meaning that $\frac{\partial }{\partial x}$ 
and $\frac{\partial }{\partial y}$ are retained. However, we 
neglect subsidence by assuming $\overline{w} = 0$, since 
$\overline{w}$ is typically much smaller than the horizontal 
wind components $\overline{u}$ and $\overline{v}$ in the free 
atmosphere \citep[e.g.,][]{lilly1983,schumann2019}. Large eddy simulation analysis of an upper-level turbulence event further supports this assumption, showing that mean vertical velocity remains small in the stably stratified free atmosphere (Figure~6.5 in \citealt{rogel2023}), and therefore, the contribution of its 
horizontal gradients to shear production is insignificant.

Under these assumptions, the flux Richardson number in \eqref{eq:Rif} reduces to

\begin{equation}
\begin{aligned}
Ri_f =
\frac{\tfrac{g}{\overline{\theta}}\,\overline{w'\theta'}}
{
\left(\overline{u'w'}\,\frac{\partial \overline{u}}{\partial z}
+ \overline{v'w'}\,\frac{\partial \overline{v}}{\partial z}
+ \overline{u'u'}\,\frac{\partial \overline{u}}{\partial x}
+ \overline{u'v'}\,\frac{\partial \overline{u}}{\partial y}
+ \overline{v'u'}\,\frac{\partial \overline{v}}{\partial x}
+ \overline{v'v'}\,\frac{\partial \overline{v}}{\partial y}
\right)
}
\end{aligned}
\label{eq:13}
\end{equation}

This expression differs from the classical flux Richardson number in that it retains four additional horizontal-shear production terms, in addition to the two classical vertical-shear terms.

The Richardson number in the form \eqref{eq:13} cannot be used operationally because it involves turbulent momentum and heat fluxes, which are not directly known. To close the problem, we apply a first-order turbulence closure (a linear eddy-viscosity approximation) and assume that the turbulent fluxes are proportional to the mean-flow gradients as follows \citep[e.g.,][]{lilly1962,smagorinsky1963,wyngaard2010}:

\begin{equation}
\begin{aligned}
\overline{w'\theta'} &= -K_H \frac{\partial \overline{\theta}}{\partial z}, \quad   
\overline{u'w'} = -K_{mv}\frac{\partial \overline{u}}{\partial z}, \quad 
\overline{v'w'} = -K_{mv}\frac{\partial \overline{v}}{\partial z}, \\
\overline{u'v'} &= \overline{v'u'} = -K_{mh}\left( 
      \frac{\partial \overline{u}}{\partial y}
    + \frac{\partial \overline{v}}{\partial x}
\right), \quad
\overline{u'u'} = -2K_{mh}\frac{\partial \overline{u}}{\partial x}, \quad
\overline{v'v'} = -2K_{mh}\frac{\partial \overline{v}}{\partial y}
\end{aligned}
\label{eq:14}
\end{equation}

$K_H$ is the heat diffusivity, $K_{mv}$ and $K_{mh}$ are the vertical and horizontal eddy viscosities, respectively.

Substituting \eqref{eq:14} into \eqref{eq:13} yields
\begin{equation}
Ri_f
=
\frac{
    -K_H \,\frac{g}{\overline{\theta}}\,\frac{\partial \overline{\theta}}{\partial z}
}{
    -K_{mv}\left( \frac{\partial \overline{u}}{\partial z} \right)^{2}
    -K_{mv}\left( \frac{\partial \overline{v}}{\partial z} \right)^{2}
    -2K_{mh}\left[ 
        \left( \frac{\partial \overline{u}}{\partial x} \right)^{2}
        + \left( \frac{\partial \overline{v}}{\partial y} \right)^{2}
    \right]
    -K_{mh}\left( 
        \frac{\partial \overline{u}}{\partial y}
        + \frac{\partial \overline{v}}{\partial x}
    \right)^{2}
}
\label{eq:15}
\end{equation}

The term 
\(
\left( \tfrac{\partial \overline{u}}{\partial x} \right)^{2}
+
\left( \tfrac{\partial \overline{v}}{\partial y} \right)^{2}
\)
can be rewritten as
\[
\frac{1}{2}
\left[
\left( 
\frac{\partial \overline{u}}{\partial x}
+
\frac{\partial \overline{v}}{\partial y}
\right)^{2}
+
\left( 
\frac{\partial \overline{u}}{\partial x}
-
\frac{\partial \overline{v}}{\partial y}
\right)^{2}
\right].
\]

Then \eqref{eq:15}  can be expressed as

\begin{equation}
Ri_f
=
\frac{
    -K_H \,\frac{g}{\overline{\theta}}\,\frac{\partial \overline{\theta}}{\partial z}
}{
    -K_{mv}\left( \frac{\partial \overline{u}}{\partial z} \right)^{2}
    -K_{mv}\left( \frac{\partial \overline{v}}{\partial z} \right)^{2}
    -K_{mh}
    \left[
        \left(
            \frac{\partial \overline{u}}{\partial x}
            +
            \frac{\partial \overline{v}}{\partial y}
        \right)^{2}
        +
        \left(
            \frac{\partial \overline{u}}{\partial x}
            -
            \frac{\partial \overline{v}}{\partial y}
        \right)^{2}
        +
        \left(
            \frac{\partial \overline{u}}{\partial y}
            +
            \frac{\partial \overline{v}}{\partial x}
        \right)^{2}
    \right]
}
\label{eq:16}
\end{equation}

Dividing \eqref{eq:16} by $K_{mv}$ gives

\begin{equation}
Ri_f
=
\frac{1}{Pr_t}
\,
\frac{
    \displaystyle \frac{g}{\overline{\theta}} \frac{\partial \overline{\theta}}{\partial z}
}{
    \displaystyle
    \left( \frac{\partial \overline{u}}{\partial z} \right)^{2}
    +
    \left( \frac{\partial \overline{v}}{\partial z} \right)^{2}
    +
    \frac{K_{mh}}{K_{mv}}
    \left[
        \left(
            \frac{\partial \overline{u}}{\partial x}
            +
            \frac{\partial \overline{v}}{\partial y}
        \right)^{2}
        +
        \left(
            \frac{\partial \overline{u}}{\partial x}
            -
            \frac{\partial \overline{v}}{\partial y}
        \right)^{2}
        +
        \left(
            \frac{\partial \overline{u}}{\partial y}
            +
            \frac{\partial \overline{v}}{\partial x}
        \right)^{2}
    \right]
}
\label{eq:17}
\end{equation}

where \(Pr_t = K_{mv}/K_H\) is the turbulent Prandtl number.

Equation~\eqref{eq:17} therefore implies
\(
Ri_f = \tfrac{1}{Pr_t}\,Ri_g,
\)
where \(Ri_g\) is the generalized gradient Richardson number:
\begin{equation}
Ri_g
=
\frac{
    N^{2}
}{
    S_{v}^{2}
    +
    \dfrac{K_{mh}}{K_{mv}}
    \left(
        Div^{2}
        +
        D_{ST}^{2}
        +
        D_{SH}^{2}
    \right)
}
\label{eq:18}
\end{equation}

The terms are defined as follows:
\begin{align}
N^{2} &= \frac{g}{\overline{\theta}}\,\frac{\partial \overline{\theta}}{\partial z},
\qquad\text{square of the Brunt-V\"{a}is\"{a}l\"{a} frequency (static stability)}, \\[0.5em]
S_v &= 
\sqrt{
    \left( \frac{\partial \overline{u}}{\partial z} \right)^{2}
    +
    \left( \frac{\partial \overline{v}}{\partial z} \right)^{2}
},
\qquad\text{vertical shear of horizontal wind}, \\[0.5em]
Div &= 
\frac{\partial \overline{u}}{\partial x}
+
\frac{\partial \overline{v}}{\partial y},
\qquad\text{horizontal divergence}, \\[0.5em]
D_{ST} &= 
\frac{\partial \overline{u}}{\partial x}
-
\frac{\partial \overline{v}}{\partial y},
\qquad\text{stretching deformation}, \\[0.5em]
D_{SH} &= 
\frac{\partial \overline{v}}{\partial x}
+
\frac{\partial \overline{u}}{\partial y},
\qquad\text{shearing deformation}.
\end{align}

The ratio \(K_{mh}/K_{mv}\) represents the anisotropy of turbulent mixing, i.e., the relative magnitude of the horizontal to vertical eddy viscosities.

Equation~\eqref{eq:18} can also be written as
\begin{equation}
Ri_g
=
\frac{
    N^{2}
}{
    S_{v}^{2}
    +
    \dfrac{K_{mh}}{K_{mv}}
    \left(
        Div^{2}
        +
        DEF^{2}
    \right)
}
\label{eq:25}
\end{equation}

where \(DEF = \sqrt{D_{SH}^{2} + D_{ST}^{2}}\) is the total horizontal deformation.

We note that the first-order turbulence closure used in this derivation follows the approach introduced by \citet{lilly1962} and \citet{smagorinsky1963}, and that the WRF model includes 
an anisotropic turbulence parameterization with $K_{mh} \neq K_{mv}$ \citep{skamarock2019description}, consistent with the physical concept underlying $Ri_{\mathrm{new}}$.

We hereafter refer to the expression in Equation~\eqref{eq:25} as the new Richardson number, \(Ri_{\mathrm{new}}\). In the following sections, \(Ri_{\mathrm{new}}\) is evaluated together with the traditional gradient Richardson number \(Ri_{\mathrm{old}}\) and a well-established turbulence index (TI1) to assess its diagnostic skill for turbulence in the upper troposphere and lower stratosphere.

\section{Data and Methods}
\subsection{Turbulence diagnostics}

In addition to the new Richardson number $Ri_{\text{new}}$ (Equation~\eqref{eq:25}), we compute two commonly used turbulence diagnostics for comparison:  
(1) the traditional gradient Richardson number ($Ri$; \citealp{richardson1922}), and  
(2) the Turbulence Index~1 (TI1) of \citet{Ellrod1992}.

The traditional Richardson number, hereafter denoted $Ri_{\text{old}}$, is defined as
\begin{equation}
Ri_{\text{old}} = \frac{N^{2}}{S_{v}^{2}},
\end{equation}
which corresponds to the special case of $Ri_{\text{new}}$ with \(K_{mh}/K_{mv} = 0\).

In gridded atmospheric fields, the gradient Richardson number is computed as a bulk Richardson number, with spatial derivatives approximated using finite differences. Small values of $Ri$ indicate that shear production is large compared to static stability, creating favorable conditions for the development of Kelvin–Helmholtz instability and clear-air turbulence \citep[e.g.,][]{Dutton}. This interpretation applies to both $Ri_{\text{old}}$ and $Ri_{\text{new}}$, with $Ri_{\text{new}}$ additionally incorporating the effects of horizontal deformation and divergence. Thus, comparing $Ri_{\text{old}}$ and $Ri_{\text{new}}$ allows us to assess the added value of including horizontal wind shear in the Richardson number.

The TI1 diagnostic is defined as
\begin{equation}
TI1 = S_{v}DEF,
\end{equation}

Large TI1 values indicate a higher likelihood of turbulence occurrence. TI1 is a  semi-empirical diagnostic that is widely used in operational forecasting and research studies because it is considered one of the best-performing indicators of upper-level turbulence \citep{sharman2017}. For this reason, we include TI1 for comparison with $Ri_{\text{new}}$.

\subsection{ERA5 reanalysis}

To compute the turbulence diagnostics \(Ri_{\mathrm{new}}\), \(Ri_{\mathrm{old}}\), and  $\mathrm{TI1}$ index, we use atmospheric fields from the ERA5 reanalysis \citep{ERA5}, produced by the European Centre for Medium-Range Weather Forecasts (ECMWF). ERA5 provides hourly global data on a regular 0.25° grid ($\sim$31~km) with 137 hybrid sigma–pressure levels. The variables used in this study include the zonal and meridional wind components, temperature, surface pressure, and the geopotential.

These fields are used to compute the pressure level, altitude, static stability $N^2$, and vertical and horizontal wind shears, from which turbulence diagnostics are derived. Spatial gradients are calculated using second-order centered finite differences. To enable direct comparison with \textit{in situ} aircraft turbulence observations, the ERA5-derived diagnostics are interpolated in the horizontal, vertical, and time using linear interpolation.

\subsection{EDR turbulence reports}
To evaluate the performance of the turbulence diagnostics, we use automated turbulence reports from commercial aircraft available through the Meteorological Assimilation Data Ingest System (MADIS) database maintained by the National Oceanic and Atmospheric Administration (NOAA).  
These reports are based on the Aircraft Communications Addressing and Reporting System (ACARS) and provide estimates of the cube root of the eddy dissipation rate, $\mathrm{EDR} = \epsilon^{1/3}$, in units of m$^{2/3}$\,s$^{-1}$. EDR has been adopted by the International Civil Aviation Organization as the standard metric for atmospheric turbulence intensity \citep{ICAO2001}. It is particularly useful because it provides an aircraft-independent measure of turbulence intensity \citep{sharman2014}. EDR is computed from spectral analysis of the vertical wind under the assumption of homogeneous and isotropic turbulence at inertial subrange scales of $\sim$10~m to 1~km \citep{sharman2014,cornman2016}. On-board algorithms derive EDR estimates every minute from 10-s Fourier transforms of the vertical wind measured at 8~Hz \citep{kaluza2025}. Energy injected by shear production at the resolved scales---whether vertical or 
horizontal in origin---is transferred to smaller scales 
where it is dissipated. EDR therefore reflects the total 
energy transfer across scales and can be used to evaluate 
diagnostics that include both vertical and horizontal 
shear contributions. The dataset and preprocessing procedures are described in detail in \cite{kaluza2025}. In brief, we exclude reports below 8 km altitude as well as those flagged as questionable or rejected by the MADIS internal consistency checks. Only regularly sampled 1-minute reports along flight tracks are retained \citep{kaluza2025}.
Turbulent intensities are classified using three different EDR thresholds:  
\begin{equation}
\begin{aligned}
\text{light-or-greater (LOG):}   &\ \mathrm{EDR} \ge 0.10\ \mathrm{m}^{2/3}\mathrm{s}^{-1}, \\
\text{moderate-or-greater (MOG):}&\ \mathrm{EDR} \ge 0.20\ \mathrm{m}^{2/3}\mathrm{s}^{-1}, \\
\text{severe-or-greater (SOG):}  &\ \mathrm{EDR} \ge 0.30\ \mathrm{m}^{2/3}\mathrm{s}^{-1}.
\end{aligned}
\end{equation}

\begin{figure}[t]
    \centering
    \includegraphics[width=\textwidth]{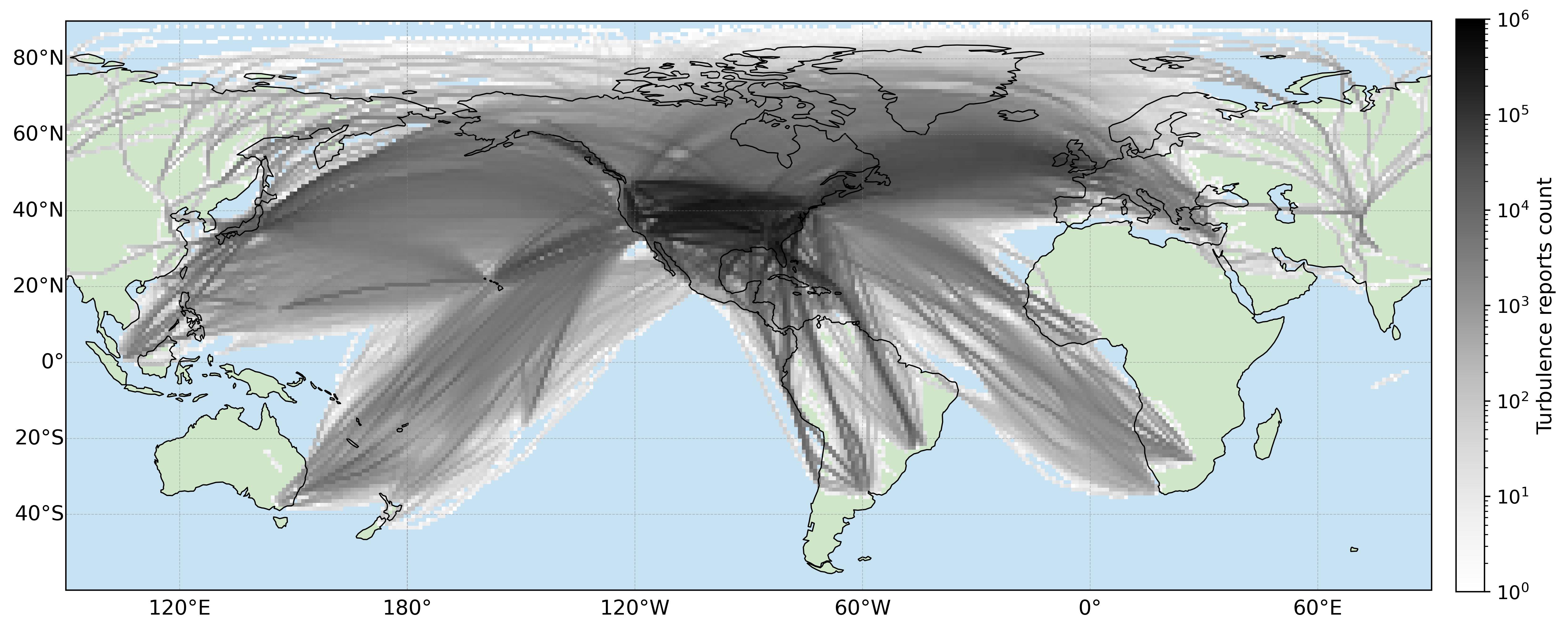}
    \caption{Spatial distribution of ACARS turbulence reports from the NOAA MADIS archive for the period 1 January 2017 to 31 December 2024. Shading indicates the number of reports on a logarithmic scale, binned onto a 1°~×~1° latitude–longitude grid. Only reports above 8~km flight altitude are included. The total number of observations is 247 730 014.}

    \label{fig:fig1}
\end{figure}



\subsection{Verification metrics}
\label{sec:metrics}

The performance of each turbulence diagnostic is assessed using categorical verification metrics derived from a $2\times2$ contingency table (Table~\ref{tab:tab1}; \citealp{gill2016}), which compares predicted and observed turbulence events based on EDR measurements from MADIS ACARS reports. For each diagnostic ($Ri$, $Ri_{\mathrm{new}}$, and the $\mathrm{TI1}$ index), a binary classification—turbulent or non-turbulent—is obtained by applying an EDR threshold, for example EDR~$\ge$~0.2~m$^{2/3}$~s$^{-1}$, commonly used to identify moderate-or-greater (MOG) turbulence. Each collocated sample is then classified into one of four contingency-table categories depending on whether turbulence was observed and/or predicted.

\begin{table}[h!]
\centering
\caption{Contingency table used to evaluate the performance of the turbulence diagnostics. Each collocated model–observation pair is classified according to whether turbulence was observed (EDR~$>$~threshold) and/or predicted by a given diagnostic ($Ri$, $Ri_\mathrm{new}$,  $\mathrm{TI1}$). The four contingency-table categories are used to compute standard verification metrics such as the probability of detection (POD; equation \eqref{eq:POD-POFD}), probability of false detection (POFD; equation \eqref{eq:POD-POFD}), and the true skill statistic (TSS; equation \eqref{eq:TSS}).}
\label{tab:tab1}
\vspace{2mm}
\begin{tabular}{lcc}
\hline
\hline
 & Observed turbulence & No observed turbulence \\
\hline
Predicted turbulence & True positive (TP) & False positive (FP) \\
No predicted turbulence & False negative (FN) & True negative (TN) \\
\hline
\end{tabular}
\label{tab:contingency}
\end{table}

From Table~\ref{tab:contingency}, several performance measures are derived. The probability of detection (POD, or true positive rate) quantifies the fraction of observed turbulence events correctly identified by the diagnostic, whereas the probability of false detection (POFD, or false positive rate) represents the fraction of non-turbulent cases incorrectly classified as turbulent. A perfect score for POD is 1 and the worst score is 0, while for POFD a perfect score is 0 and the worst is 1.

\begin{align}
\mathrm{POD} &= \frac{\mathrm{TP}}{\mathrm{TP} + \mathrm{FN}}, &
\mathrm{POFD} &= \frac{\mathrm{FP}}{\mathrm{FP} + \mathrm{TN}}.
\label{eq:POD-POFD} 
\end{align}

To evaluate performance across all possible thresholds, we construct a sequence of $2\times2$ contingency tables by progressively varying the turbulence threshold applied to each diagnostic. Plotting the corresponding POD against the POFD produces a Receiver Operating Characteristic (ROC) curve, which summarizes the trade-off between hits and false alarms across thresholds and provides a visual measure of diagnostic skill \citep[e.g.,][]{sharman2006, gill2016, sharman2017}. 
A commonly used performance metric is the area under the ROC curve (AUC), a single quantitative measure of discrimination ranging from 0.5 (random performance) to 1.0 (perfect skill) \citep{gill2016, sharman2016}. However, AUC values should be interpreted with care, as they can depend on event frequency and sampling characteristics \citep{kaluza2025}.  

In addition to ROC analysis, we compute the True Skill Statistic (TSS), also known as the Hanssen–Kuipers discriminant or Peirce’s Skill Score, which combines the POD and the POFD:
\begin{equation}
\mathrm{TSS} = \mathrm{POD} - \mathrm{POFD}
             = \frac{\mathrm{TP}}{\mathrm{TP} + \mathrm{FN}} - \frac{\mathrm{FP}}{\mathrm{FP} + \mathrm{TN}}.
\label{eq:TSS}             
\end{equation}
TSS ranges from $-1$ to $+1$, with 0 indicating no skill and $+1$ perfect discrimination. It is a robust measure that balances detection and false alarms, making it particularly interesting for rare-event verification, such as turbulence \citep[e.g.,][]{gill2016,sharman2017}. In this study, TSS is also used to determine the optimal threshold that maximizes diagnostic performance.

Together, these complementary metrics provide a comprehensive assessment of diagnostic performance and discrimination skill. They are computed for all diagnostics—the classical Richardson number ($Ri$), the new formulation ($Ri_\mathrm{new}$), and the turbulence index $\mathrm{TI1}$—using the MADIS ACARS turbulence reports from 2017 to 2024, allowing for a robust evaluation of the new Richardson number $Ri_\mathrm{new}$.

\section{Results}

\subsection{Evaluation of $Ri_{\mathrm{new}}$}

\begin{figure}[h]
    \centering
    \includegraphics[width=0.35\textwidth]{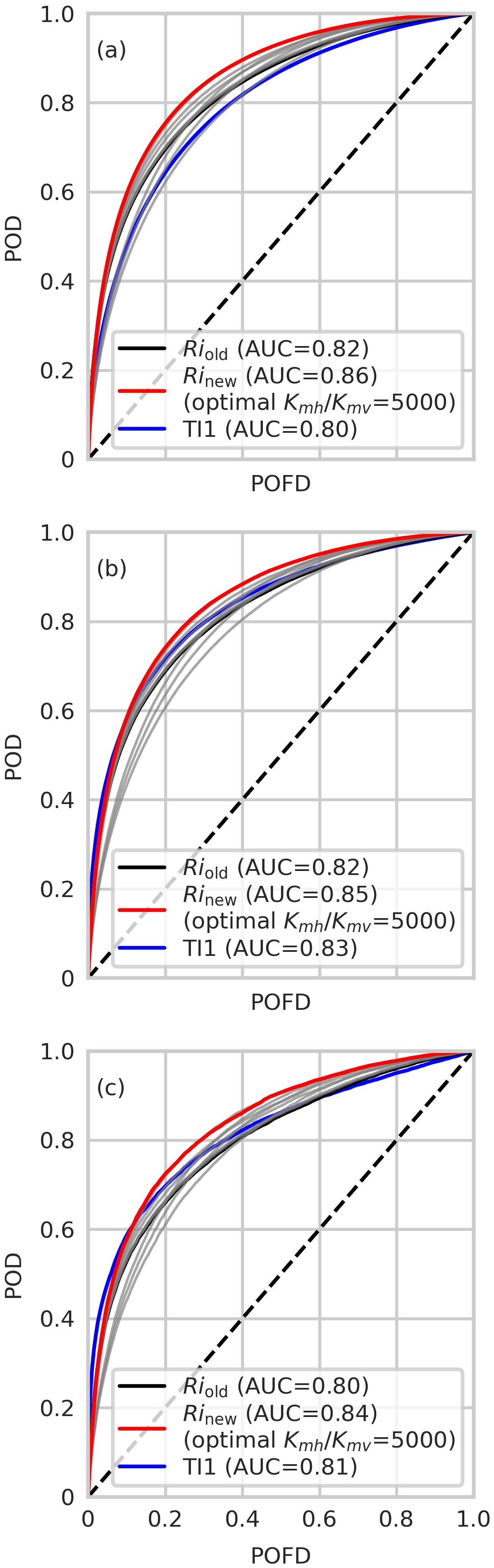}
    \caption{ROC curves for TI1 index (blue), the classical Richardson number (Ri$_{\mathrm{old}}$, black), and $Ri_{\mathrm{new}}$ with $K_{mh}/K_{mv} = 5000$ (red) for turbulence intensity thresholds of (a) EDR $=$ 0.10, (b) EDR $=$ 0.20, and (c) EDR $=$ 0.30 $\mathrm{m}^{2/3}\,\mathrm{s}^{-1}$. Gray curves show ROC curves for $Ri_{\mathrm{new}}$ computed 
using different values of $K_{mh}/K_{mv}$. The diagonal dashed line denotes no skill. Values in parentheses give the area under the ROC curve (AUC). The total number of observations used was 247\,730\,014, with 1\,770\,543 ($\approx$ 0.71\%), 192\,447 ($\approx$ 0.08\%), and 19\,705 ($\approx$ 0.01\%) turbulence events for (a) EDR $\ge$ 0.10 (LOG), (b) EDR $\ge$ 0.20 (MOG), and (c) EDR $\ge$ 0.30 $\mathrm{m}^{2/3}\,\mathrm{s}^{-1}$ (SOG), respectively.}
    
    \label{fig:fig2}
\end{figure}

Figure~\ref{fig:fig2} shows the derived ROC curves and compares the performance of the commonly used TI1 index, the classical Richardson number (Ri$_{\mathrm{old}}$), and the new formulation Ri$_{\mathrm{new}}$ against MADIS ACARS turbulence reports for three turbulence thresholds: light-or-greater (LOG, EDR $\ge$ 0.10~m$^{2/3}$\,s$^{-1}$), moderate-or-greater (MOG, EDR $\ge$ 0.20~m$^{2/3}$\,s$^{-1}$), and severe-or-greater (SOG, EDR $\ge$ 0.30~m$^{2/3}$\,s$^{-1}$). Across all thresholds, Ri$_{\mathrm{new}}$ outperforms both Ri$_{\mathrm{old}}$ and TI1, resulting in consistently higher AUC values. The grey curves illustrate the sensitivity of Ri$_{\mathrm{new}}$ to the mixing-ratio parameter $K_{mh}/K_{mv}$; although several values show improvements relative to Ri$_{\mathrm{old}}$, a ratio of 5000 gives the highest AUC across the three thresholds. This magnitude is consistent with previously reported estimates of $K_{mh}/K_{mv}$ (10$^{4}$–10$^{5}$) in strongly stratified geophysical flows, including the free troposphere and the deep ocean \citep[e.g.,][]{pisso2009,inall2013}.

For LOG turbulence (EDR~$\ge$~0.10), Ri$_{\mathrm{old}}$ performs slightly better than TI1, with AUC values of 0.82 and 0.80, respectively. Ri$_{\mathrm{new}}$ with $K_{mh}/K_{mv}=5000$ outperforms both, achieving an AUC of 0.86. For MOG (EDR~$\ge$~0.20) and SOG turbulence (EDR~$\ge$~0.30), TI1 performs better than Ri$_{\mathrm{old}}$, but Ri$_{\mathrm{new}}$ still provides the highest diagnostic skill among all diagnostics. Additional results with different turbulence thresholds commonly used in the literature (EDR~$\ge$~0.15, 0.22, 0.34, and 0.45~m$^{2/3}$\,s$^{-1}$) are shown in Figure~\ref{fig:fig1A}. In general, these results demonstrate that Ri$_{\mathrm{new}}$ provides a more skillful turbulence discriminator than both Ri$_{\mathrm{old}}$ and TI1 across the full range of turbulence intensity thresholds.

\begin{figure}[h]
    \centering
    \includegraphics[width=\textwidth]{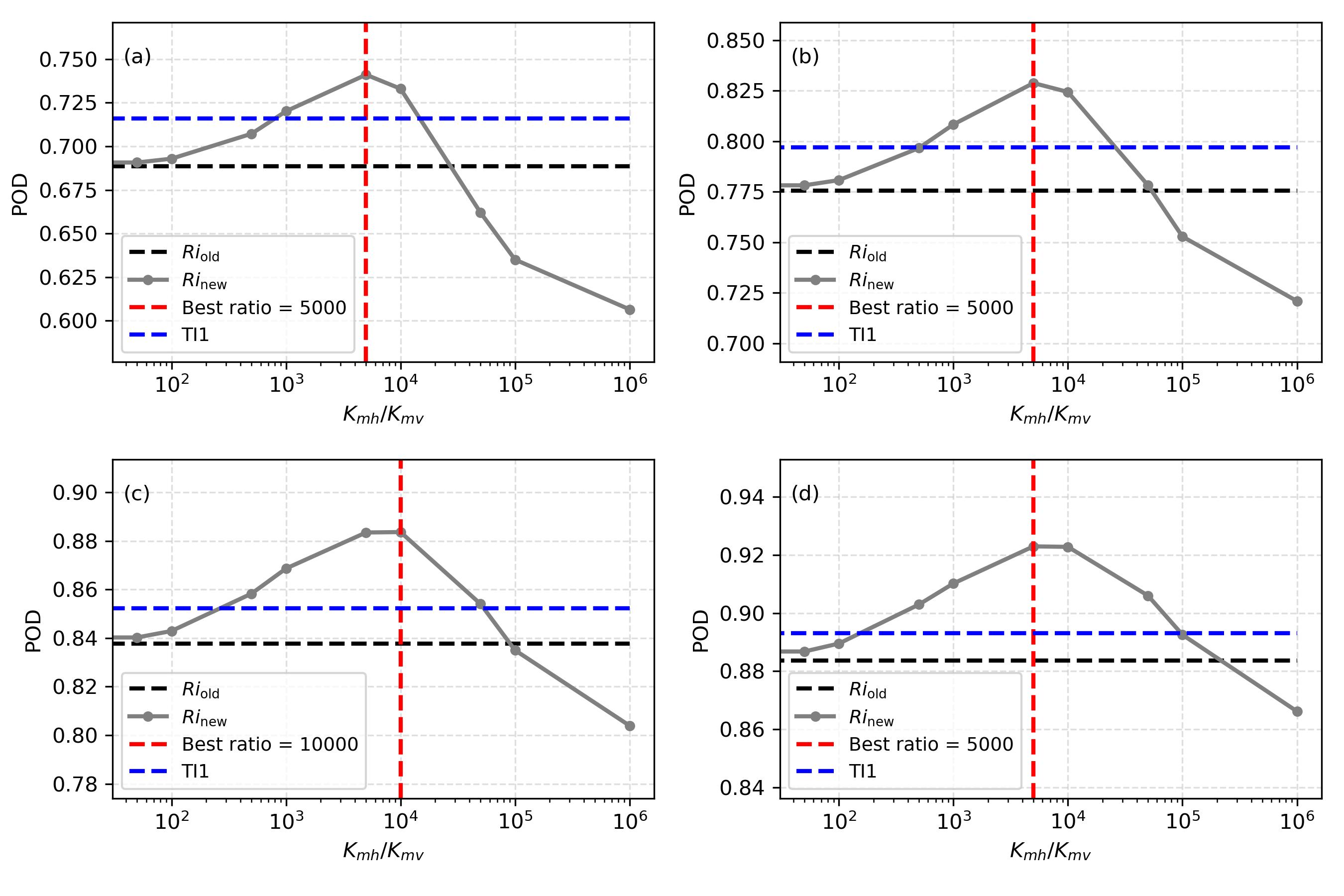}
    \caption{
Probability of Detection (POD) as a function of the ratio 
$K_{mh}/K_{mv}$ for the new Richardson number formulation 
Ri$_{\mathrm{new}}$ (solid grey) evaluated at four fixed values of the  Probability of False Detection (POFD): (a) 20\%, (b) 30\%, (c) 40\%,  and (d) 50\%. Horizontal dashed lines show the skill of TI1 (blue) and  Ri$_{\mathrm{old}}$ (black), while the vertical red dashed line shows the value of $K_{mh}/K_{mv}$ that yields the highest POD for MOG turbulence (EDR~$\ge$~0.20\,m$^{2/3}$\,s$^{-1}$).}
    \label{fig:fig3}
\end{figure}

Although ROC curves and their associated AUC values provide an integrated measure of the ability to discriminate between turbulent and non-turbulent conditions, they do not necessarily indicate how a diagnostic performs at operationally relevant false-alarm rates. In practice, aviation forecasting aims to maximize the Probability of Detection (POD) while keeping the Probability of False Detection (POFD) as low as possible. 

Figure~\ref{fig:fig3} therefore shows the PODs of TI1, Ri$_{\mathrm{old}}$, and Ri$_{\mathrm{new}}$ at lower-end POFD levels (20\%, 30\%, 40\%, and 50\%) for MOG turbulence (EDR~$\ge$~0.20\,m$^{2/3}$\,s$^{-1}$), evaluated across a range of $K_{mh}/K_{mv}$ ratios to assess sensitivity to this parameter. Across all POFD thresholds, Ri$_{\mathrm{new}}$ yields higher POD values than Ri$_{\mathrm{old}}$ for any $K_{mh}/K_{mv}$ ratio above zero, except when  the ratio becomes excessively large ($\gtrsim 5\times10^{4}$), which degrades performance. TI1 consistently outperforms Ri$_{\mathrm{old}}$, but  Ri$_{\mathrm{new}}$ with ratios in the range $10^{3}$--$10^{4}$ exceeds both. For example, at POFD = 30\%, $Ri_{\mathrm{new}}$ with $K_{mh}/K_{mv} = 5000$ achieves a POD of approximately 83\%, compared with about 77.5\% for Ri$_{\mathrm{old}}$. Although panel (c) in Figure~3 shows the highest POD at $K_{mh}/K_{mv} = 10^{4}$, the difference relative to $K_{mh}/K_{mv} = 5000$ is very small, indicating that the diagnostic skill is largely insensitive within this well-performing range. Similar results are obtained for the LOG and SOG turbulence thresholds (Figs.~\ref{fig:fig2A} and \ref{fig:fig3A}), demonstrating that the enhanced diagnostic skill of Ri$_{\mathrm{new}}$ is consistent across turbulence intensities and at operationally relevant POFD levels.

Table~\ref{tab:tab2} summarizes additional performance metrics derived from the $2\times2$ contingency tables—AUC, POD, PODN ($=1-\mathrm{POFD}$), and TSS—computed using the optimal index threshold that maximizes TSS along the ROC curve for TI1, $Ri_{\mathrm{old}}$, and $Ri_{\mathrm{new}}$ for several $K_{mh}/K_{mv}$ ratios under MOG turbulence conditions (EDR~$\ge$~0.20\,m$^{2/3}$\,s$^{-1}$). Maximizing $\mathrm{TSS} = \mathrm{POD} + \mathrm{PODN} - 1$, provides a balanced measure of performance because it simultaneously reflects the ability to detect turbulence events and limit false alarms. A perfect discriminator yields $\mathrm{TSS}=1$. The highest TSS is obtained with $Ri_{\mathrm{new}}$ ($K_{mh}/K_{mv}=5000$), reaching 0.556, compared with 0.507 for $Ri_{\mathrm{old}}$ and 0.531 for TI1.

Although the PODN of $Ri_{\mathrm{new}}$ with $K_{mh}/K_{mv} = 5000$ is comparable to that of $Ri_{\mathrm{old}}$ and slightly lower than TI1, the corresponding POD is substantially higher than for either diagnostic. Thus, the combined TSS metric more clearly reflects the improvement provided by $Ri_{\mathrm{new}}$ by quantifying the trade-off between detection and false alarms. A higher PODN could be achieved by lowering the classification turbulence threshold, but only at the cost of reducing POD, reflecting the strong class imbalance inherent in turbulence occurrence (far more non-turbulent than turbulent reports). Finally, the optimal index threshold—defined as the value that maximizes TSS—for $Ri_{\mathrm{new}}$ ($K_{mh}/K_{mv}=5000$) is 1.298, compared with 2.165 for $Ri_{\mathrm{old}}$.

To assess whether the results depend on the EDR intensity threshold, the same performance metrics were computed for LOG 
(EDR~$\ge$~0.10) and SOG (EDR~$\ge$~0.30) turbulence categories 
(Tables~A1 and~A2). Across all thresholds, $Ri_{\mathrm{new}}$ remains the best-performing diagnostic, showing the highest AUC and TSS values, with the value of $K_{mh}/K_{mv}$ that maximizes diagnostic skill consistently close to $5000$. The improvement is most pronounced for LOG turbulence and slightly lower for SOG turbulence. The relative performance of TI1 and $Ri_{\mathrm{old}}$ depends on the threshold—TI1 performs worse than $Ri_{\mathrm{old}}$ for LOG turbulence, but better for both MOG and SOG turbulence—whereas $Ri_{\mathrm{new}}$ outperforms both in all cases. It should be noted that the optimal index threshold for $Ri_{\mathrm{old}}$ and $Ri_{\mathrm{new}}$ shows little dependence on the EDR intensity threshold, in contrast to TI1, whose optimal threshold increases substantially from LOG to MOG and SOG turbulence. This is consistent with previous results showing that $Ri$ primarily discriminates between turbulent and non-turbulent conditions, whereas shear-based diagnostics such as TI1 tend to correlate more strongly with turbulence intensity \citep{Kaluza2022}.

In summary, the evaluation across multiple turbulence thresholds, false-alarm levels, and various statistical performance metrics demonstrates that $Ri_{\mathrm{new}}$ consistently outperforms $Ri_{\mathrm{old}}$ and TI1 index, with a diagnostic skill that remains robust for LOG, MOG, and SOG turbulence intensities. The following subsections present case studies comparing the spatial distribution of $Ri_{\mathrm{old}}$ and $Ri_{\mathrm{new}}$ with 
collocated turbulence observations, investigate the sensitivity of the best-performing value of $K_{mh}/K_{mv}$ to horizontal grid resolution, assess the statistical significance of the improvement, and examine the seasonal and regional variability 
in diagnostic performance.

\begin{table}[h]
\centering
\caption{Statistical performance metrics computed from the $2\times2$ contingency tables for TI1, $Ri_{\text{old}}$, and $Ri_{\text{new}}$ for MOG turbulence (EDR~$\ge$~0.20\,m$^{2/3}$\,s$^{-1}$). Metrics shown are the area under the ROC curve (AUC), the probability of detection (POD), the probability of detection of non-events (PODN $=1-\mathrm{POFD}$), the True Skill Statistic (TSS), and the optimal index threshold (i.e., the value that maximizes TSS). The total number of observations used was 247\,730\,014. The best-performing $Ri_{\text{new}}$ obtained with $K_{mh}/K_{mv}=5000$ is highlighted in bold, as it yields the highest AUC and TSS.}

\small
\begin{tabular}{lccccc}
\hline \hline
Diagnostic & AUC & POD & PODN & TSS & Threshold \\
\hline
$\mathrm{TI1}$                     & 0.834 & 0.724 & 0.806 & 0.531 & $5.40\times10^{-7}$ \\
$Ri_{\mathrm{old}}$               & 0.823 & 0.712 & 0.795 & 0.507 & 2.165 \\
$Ri_{\mathrm{new}}$ $(K_{mh}/K_{mv}=50)$      & 0.829 & 0.713 & 0.795 & 0.509 & 2.145 \\
$Ri_{\mathrm{new}}$ $(K_{mh}/K_{mv}=100)$     & 0.831 & 0.722 & 0.789 & 0.511 & 2.191 \\
$Ri_{\mathrm{new}}$ $(K_{mh}/K_{mv}=500)$     & 0.840 & 0.740 & 0.784 & 0.525 & 2.114 \\
$Ri_{\mathrm{new}}$ $(K_{mh}/K_{mv}=1000)$    & 0.846 & 0.759 & 0.778 & 0.539 & 2.032 \\
$Ri_{\mathrm{new}}$ $(K_{mh}/K_{mv}=5000)$    & \textbf{0.855} & \textbf{0.762} & \textbf{0.793} & \textbf{0.556} & \textbf{1.298} \\
$Ri_{\mathrm{new}}$ $(K_{mh}/K_{mv}=10000)$   & 0.849 & 0.778 & 0.767 & 0.546 & 1.019 \\
$Ri_{\mathrm{new}}$ $(K_{mh}/K_{mv}=50000)$   & 0.817 & 0.767 & 0.713 & 0.480 & 0.345 \\
$Ri_{\mathrm{new}}$ $(K_{mh}/K_{mv}=100000)$  & 0.802 & 0.758 & 0.694 & 0.452 & 0.234 \\
$Ri_{\mathrm{new}}$ $(K_{mh}/K_{mv}=1000000)$ & 0.782 & 0.719 & 0.692 & 0.417 & 0.025 \\
\hline
\label{tab:tab2}
\end{tabular}
\end{table}

\clearpage
\newpage

\subsection{Spatial distribution of $Ri_{\mathrm{old}}$ and $Ri_{\mathrm{new}}$: case studies}

\begin{figure}[h]
    \centering
    \includegraphics[width=\textwidth]{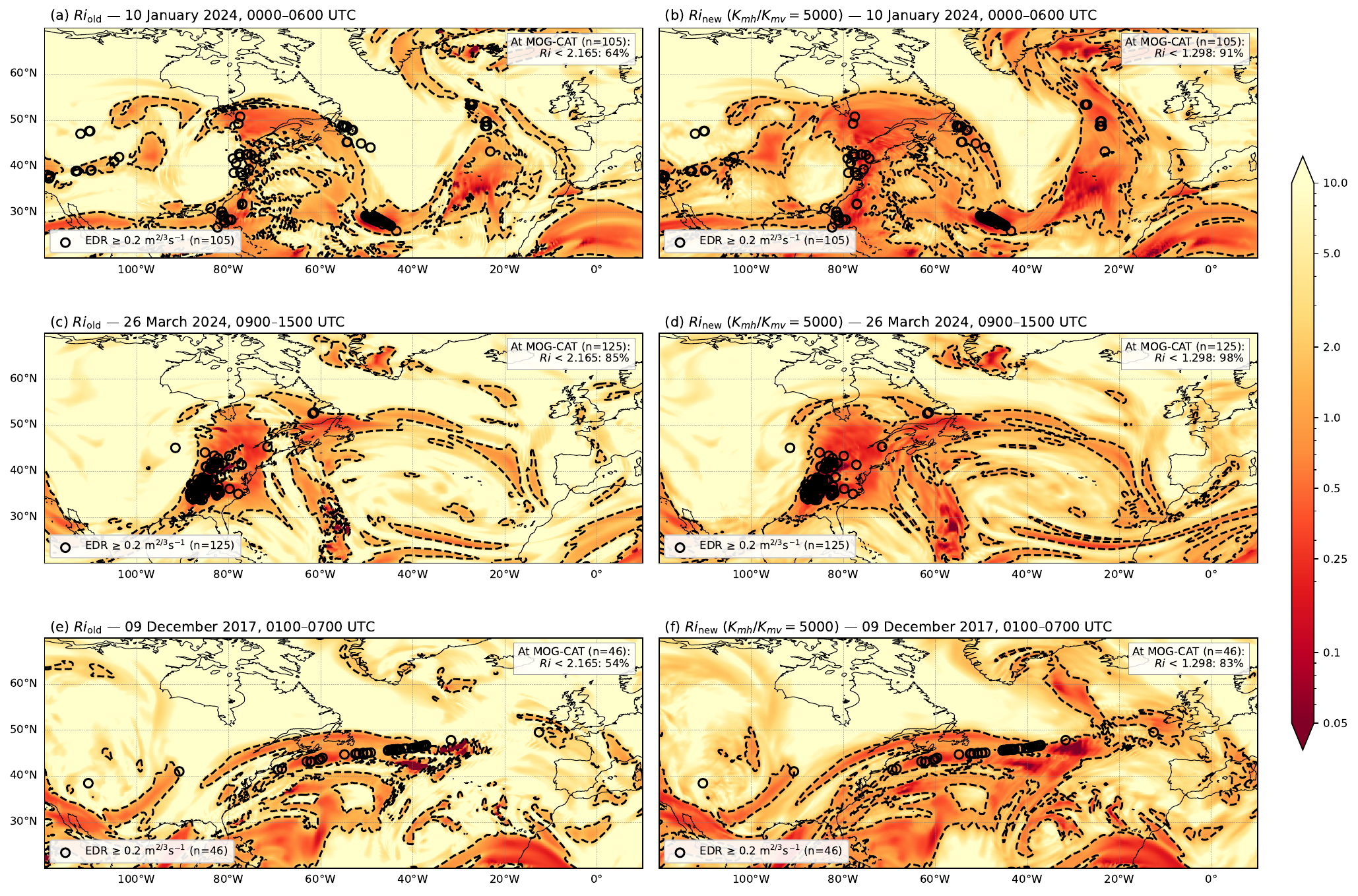}
\caption{Spatial distribution of $Ri_{\mathrm{old}}$ (left column) and $Ri_{\mathrm{new}}$ (right column, $K_{mh}/K_{mv}=5000$) computed from ERA5 at model level 78 ($\sim$241~hPa, $\sim$11~km) for three representative case study days: (a,~b) 10 January 2024 (0000--0600~UTC), (c,~d) 26 March 2024 (0900--1500~UTC), and (e,~f) 09 December 2017 (0100--0700~UTC). Each field represents a 6-hour mean centred on the period of peak MOG-CAT activity. Shading shows the Richardson number (log scale), with darker (red) colors indicating lower values (i.e., dynamically unstable conditions). Dashed contours indicate the optimal TSS detection threshold ($Ri_{\mathrm{old}} = 2.165$ in the left column; $Ri_{\mathrm{new}} = 1.298$ in the right column). Black circles indicate collocated MOG-CAT turbulence reports from MADIS ACARS (EDR~$\geq 0.2$~m$^{2/3}$\,s$^{-1}$) within the same 6-hour window at cruise altitudes ($\sim$10--12~km). The fraction of MOG-CAT events below the detection threshold is shown in the upper right of each panel.}
    \label{fig:fig4}
\end{figure}

To complement the global statistical evaluation presented in Section~4a, we examine the spatial distribution of $Ri_{\mathrm{old}}$ and $Ri_{\mathrm{new}}$ for three case study days, together with collocated MOG-CAT turbulence reports from MADIS ACARS (EDR~$\geq$~0.2~m$^{2/3}$~s$^{-1}$) within the same 
6-hour window at cruise altitudes ($\sim$10--12~km) (Figure~\ref{fig:fig4}). For each case, the diagnostic fields are 
computed as a 6-hour mean centred on the period of peak MOG-CAT 
activity at model level 78 ($\sim$241~hPa, $\sim$11~km). This 
averaging window is chosen to maximize the number of collocated 
EDR observations while remaining representative of the synoptic-scale flow. The spatial structures of both $Ri_{\mathrm{old}}$ and 
$Ri_{\mathrm{new}}$ vary only marginally within this 6-hour window 
(not shown). The three cases span different seasons and large-scale flow conditions: 10~January~2024 (winter, 0000--0600~UTC, $n=105$), 26~March~2024 (spring, 0900--1500~UTC, $n=125$), and 
09~December~2017 (winter, 0100--0700~UTC, $n=46$).

\begin{figure}[h]
    \centering
    \includegraphics[width=\textwidth]{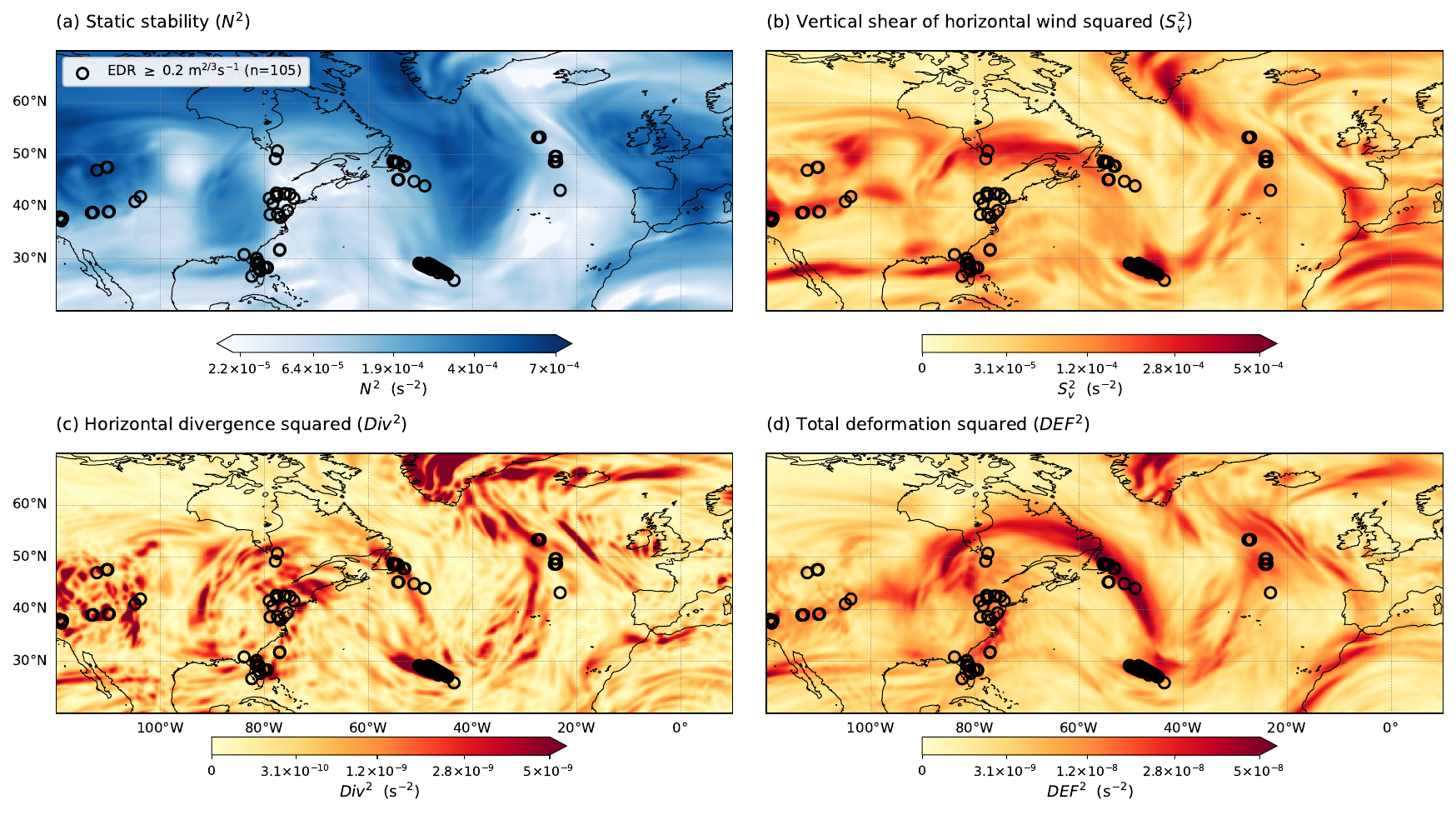}
\caption{Spatial distribution of the four ERA5 diagnostic fields used in the $Ri_{\mathrm{new}}$ formulation at model level 78 
($\sim$241~hPa, $\sim$11~km) for 10 January 2024 (0000--0600~UTC, 6-hour mean): (a) static stability ($N^{2}$), (b) vertical shear of horizontal wind squared ($S_{v}^{2}$), (c) horizontal divergence squared ($Div^{2}$), and (d) total deformation squared ($DEF^{2}$). Note the different colour bar scales across panels. Black circles indicate collocated MOG-CAT turbulence reports from MADIS ACARS (EDR~$\geq 0.2$~m$^{2/3}$\,s$^{-1}$) within the same 6-hour window at cruise altitudes ($\sim$10--12~km).}
    \label{fig:fig5}
\end{figure}

In all three cases, $Ri_{\mathrm{new}}$ exhibits broadly similar spatial patterns to $Ri_{\mathrm{old}}$, with low values concentrated along the jet stream and frontal regions where 
vertical wind shear is strong. However, $Ri_{\mathrm{new}}$ identifies additional regions of dynamic instability associated with horizontal divergence and deformation that are not captured by $Ri_{\mathrm{old}}$. Figure~\ref{fig:fig5} shows the spatial distribution of the four diagnostic fields used in $Ri_{\mathrm{new}}$ for the 10~January~2024 case. The static stability ($N^{2}$, panel a) is positive throughout the domain, consistent with stable stratification at upper-tropospheric levels. The vertical shear of horizontal wind squared ($S_{v}^{2}$, panel b) shows intense structures along the North Atlantic jet stream. The horizontal divergence squared ($Div^{2}$, panel c) and total deformation squared ($DEF^{2}$, panel d) reveal additional turbulence-prone structures in frontal zones and jet-exit regions that are not captured by $S_{v}^{2}$ alone. Several observed MOG-CAT events are located in regions of strong horizontal divergence and deformation but relatively weak vertical wind shear and strongly stable stratification (Fig.~\ref{fig:fig5}, panels a--d), suggesting that these horizontal dynamical processes play a role in triggering turbulence that $Ri_{\mathrm{old}}$ fails to capture. Despite being 4--5 orders of magnitude smaller than $S_{v}^{2}$, these horizontal terms contribute significantly to the denominator of $Ri_{\mathrm{new}}$ through the anisotropy ratio $K_{mh}/K_{mv}=5000$, which scales them to a magnitude comparable to $S_{v}^{2}$. Similar results are found for the two other case study days, as shown in the Appendix (Figs.~\ref{fig:fig4A} and~\ref{fig:fig5A}).

As a result, $Ri_{\mathrm{new}}$ captures a substantially larger fraction of observed MOG-CAT events below the optimal TSS diagnostic threshold. It is worth noting that the optimal thresholds differ between the two diagnostics: $Ri_{\mathrm{old}} = 2.165$ and $Ri_{\mathrm{new}} = 1.298$ (Table~2), reflecting the fact that the additional horizontal shear terms in the denominator of $Ri_{\mathrm{new}}$ systematically reduce its values relative to $Ri_{\mathrm{old}}$. Despite this lower threshold ---  approximately a factor of~2 smaller --- $Ri_{\mathrm{new}}$ still identifies a substantially larger fraction of MOG-CAT events: 91\% versus 64\% on 10~January~2024; 98\% versus 85\% on 26~March~2024; and 83\% versus 54\% on 09~December~2017 (Figure~\ref{fig:fig4}). This confirms that the improvement is not simply a consequence of using a lower threshold, but reflects a better spatial correspondence between $Ri_{\mathrm{new}}$ and observed turbulence locations. One might expect that the broader low-Ri regions identified by $Ri_{\mathrm{new}}$ would lead to a substantially higher false alarm rate. However, Table~2 shows that the probability of false detection increases only marginally from 0.205 ($Ri_{\mathrm{old}}$) to 0.207 ($Ri_{\mathrm{new}}$, $K_{mh}/K_{mv}=5000$), while the probability of detection increases from 0.712 to 0.762. This confirms that $Ri_{\mathrm{new}}$ improves turbulence detection without introducing a substantial increase in false alarms, reflecting a more physically realistic representation of turbulence-prone regions. Finally, it should be noted that the MADIS ACARS observations used here are predominantly from US commercial airlines, with limited coverage from non-US airlines. Furthermore, some flights may have deviated from their planned routes to avoid known turbulence regions. The spatial distribution of MOG-CAT reports in Figure~\ref{fig:fig4} therefore reflects flight track coverage rather than the true spatial distribution of turbulence, and regions with low circle density should not be interpreted as turbulence-free.

\clearpage
\newpage
\subsection{Sensitivity studies}
The first sensitivity study examines whether the value of $K_{mh}/K_{mv}$ that yields the highest diagnostic skill depends on the horizontal grid resolution. In the previous section, we showed that $Ri_{\mathrm{new}}$ with any value of $K_{mh}/K_{mv}$ greater than zero yields better performance relative to $Ri_{\mathrm{old}}$, except when the ratio becomes excessively large ($\gtrsim 5\times10^{4}$). We also found that $Ri_{\mathrm{new}}$ with ratios in the range $10^{3}$--$10^{4}$ substantially exceeds the performance of TI1, and peak skill is obtained for $K_{mh}/K_{mv}\approx 5000$. These results were derived using ERA5 at its native 0.25$^\circ$ ($\sim$31~km) resolution, which is comparable to the grid spacing of operational NWP models used for aviation turbulence forecasting. 
However, the value of $K_{mh}/K_{mv}$ that yields the highest diagnostic skill may depend on horizontal resolution, because the horizontal spatial gradients appearing in the divergence and deformation terms in Equation~\eqref{eq:18} may strengthen at finer resolution and weaken at coarser resolution.

\begin{table}[h]
\centering
\caption{Value of $K_{mh}/K_{mv}$ that maximizes diagnostic 
skill and corresponding AUC for $Ri_{\mathrm{new}}$ computed using ERA5 at its native $0.25^\circ$ resolution and after regridding to $1^\circ$ and $2^\circ$. Results are shown for three turbulence intensity thresholds: LOG (EDR $\ge 0.10$), MOG (EDR $\ge 0.20$), and SOG (EDR $\ge 0.30$).}

\small
\begin{tabular}{lccc}
\hline\hline
 & ERA5 native (0.25$^\circ$) & ERA5 regridded (1$^\circ$) 
 & ERA5 regridded (2$^\circ$) \\
\hline
LOG (EDR $\ge 0.10$) 
 & 7000 (AUC = 0.860) 
 & 9000 (AUC = 0.852) 
 & 9000 (AUC = 0.848) \\[1mm]

MOG (EDR $\ge 0.20$) 
 & 5000 (AUC = 0.855) 
 & 7000 (AUC = 0.846) 
 & 7000 (AUC = 0.843) \\[1mm]

SOG (EDR $\ge 0.30$) 
 & 5000 (AUC = 0.845) 
 & 8000 (AUC = 0.831) 
 & 8000 (AUC = 0.830) \\
\hline
\label{tab:tab3}
\end{tabular}
\end{table}

Table~\ref{tab:tab3} shows the values of $K_{mh}/K_{mv}$ that maximize diagnostic skill and the corresponding AUC obtained when $Ri_{\mathrm{new}}$ is computed using ERA5 at its native $0.25^\circ$ resolution and after regridding to $1^\circ$ and $2^\circ$ grids. The range of tested values $K_{mh}/K_{mv}$ in this sensitivity experiment is slightly wider than in Table~2, which explains why the value for LOG turbulence at native resolution is 7000 rather than 5000. However, the differences in AUC are very small, consistent with the weak sensitivity within the well-performing interval identified earlier. For LOG and MOG turbulence, AUC values decrease by about 0.01 when regridding from 0.25$^\circ$ to 1$^\circ$, and by roughly 0.015 for SOG turbulence. The difference between the 1$^\circ$ and 2$^\circ$ results is minimal  ($\lesssim$0.003\% in AUC), indicating that the diagnostic is only weakly sensitive to horizontal grid spacing at these resolutions. The regridding was performed using area-averaging; an alternative approach based on subsampling the grid (i.e., retaining every fourth or eighth grid point) showed nearly identical values, confirming that the results are also weakly sensitive to the regridding method.

The best-performing value of $K_{mh}/K_{mv}$ increases slightly as 
the grid is coarsened---from values around 5000--7000 at 
$0.25^\circ$ to approximately 7000--9000 at $1^\circ$ and 
$2^\circ$ (and similarly for $3^\circ$, not shown). This is physically consistent because horizontal averaging weakens the divergence and deformation terms in Equation~\eqref{eq:18}, requiring a somewhat larger $K_{mh}/K_{mv}$ to maintain a correct balance between vertical and horizontal shear effects. It should be noted that the AUC varies by only about $\pm 0.5\%$ across the entire range of well-performing ratios ($5\times10^{3}$--$10^{4}$), indicating that the tuning of $K_{mh}/K_{mv}$ is not critical within this interval and that any value in this range yields nearly identical performance and large improvements relative to $Ri_{\mathrm{old}}$.

\begin{figure}[h]
    \centering
    \includegraphics[width=0.5\textwidth]{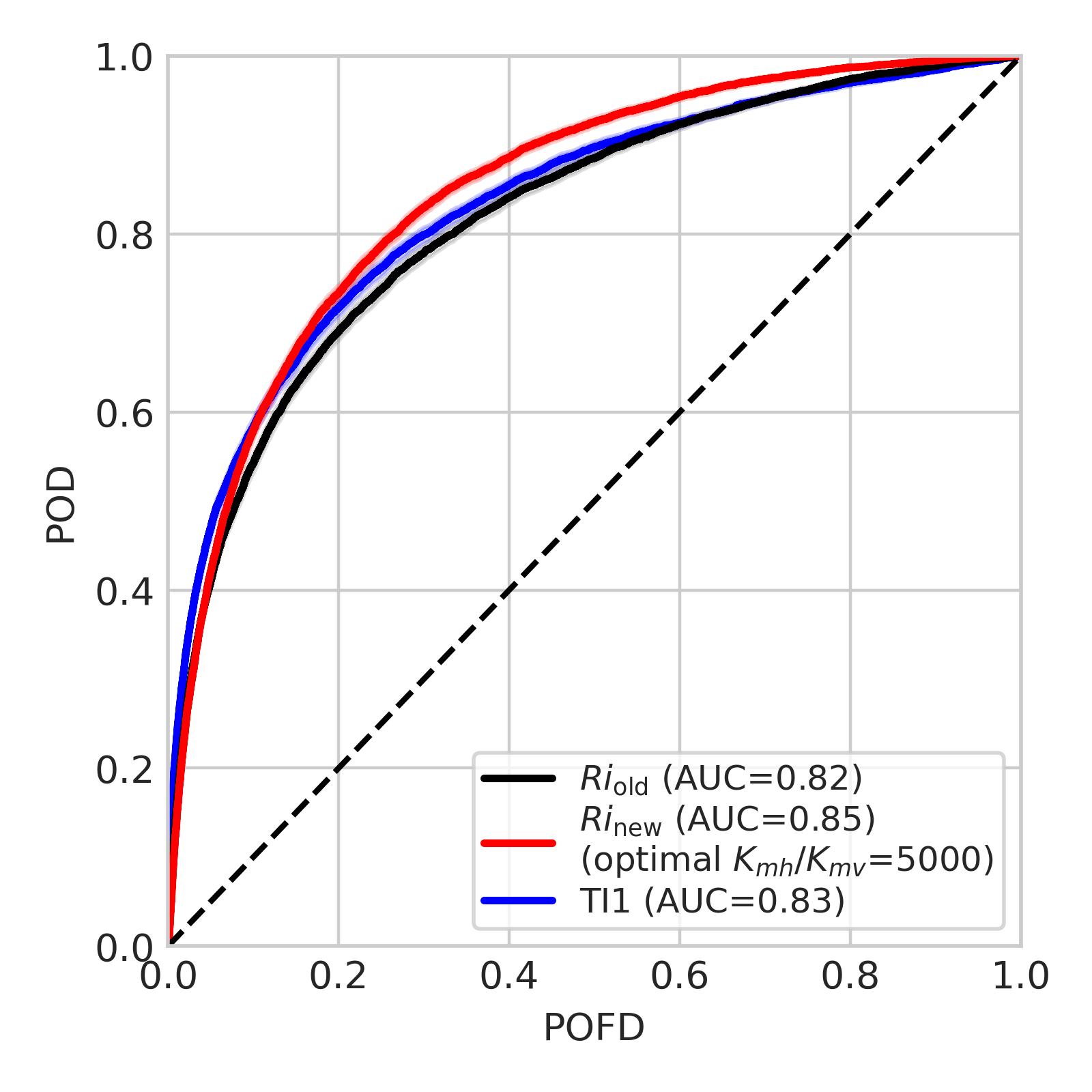}
\caption{ROC curves and 95\% confidence intervals for TI1, 
$Ri_{\mathrm{old}}$, and $Ri_{\mathrm{new}}$ (with $K_{mh}/K_{mv}=5000$), computed from 10\,000\,000 randomly sampled reports with 1000 bootstrap replications (EDR $\ge 0.20$~m$^{2/3}$\,s$^{-1}$). The uncertainty bands are smaller than the line thickness (approximately $\pm 1\%$) and are therefore almost imperceptible.}
    \label{fig:fig6}
\end{figure}

\begin{table}[h]
\centering
\caption{Bootstrap AUC calculation (mean and 95\% confidence intervals) for TI1, $Ri_{\mathrm{old}}$, and $Ri_{\mathrm{new}}$ using EDR $\ge 0.20$~m$^{2/3}$\,s$^{-1}$, a subsample of 10\,000\,000 reports, and 1000 bootstrap replications. The best-performing $Ri_{\mathrm{new}}$ (with $K_{mh}/K_{mv}=5000$) is highlighted in bold.}

\small
\begin{tabular}{lccc}
\hline\hline
Diagnostic & AUC (mean) & AUC (2.5\%) & AUC (97.5\%) \\
\hline
TI1 & 0.833 & 0.828 & 0.838 \\
$Ri_{\mathrm{old}}$ & 0.819 & 0.814 & 0.823 \\
$Ri_{\mathrm{new}}$ $(K_{mh}/K_{mv}=50)$ & 0.821 & 0.817 & 0.826 \\
$Ri_{\mathrm{new}}$ $(K_{mh}/K_{mv}=100)$ & 0.823 & 0.818 & 0.828 \\
$Ri_{\mathrm{new}}$ $(K_{mh}/K_{mv}=500)$ & 0.833 & 0.829 & 0.837 \\
$Ri_{\mathrm{new}}$ $(K_{mh}/K_{mv}=1000)$ & 0.839 & 0.835 & 0.844 \\
\textbf{$Ri_{\mathrm{new}}$ $(K_{mh}/K_{mv}=5000)$} & \textbf{0.848} & \textbf{0.843} & \textbf{0.852} \\
$Ri_{\mathrm{new}}$ $(K_{mh}/K_{mv}=10000)$ & 0.842 & 0.838 & 0.846 \\
$Ri_{\mathrm{new}}$ $(K_{mh}/K_{mv}=50000)$ & 0.810 & 0.805 & 0.814 \\
$Ri_{\mathrm{new}}$ $(K_{mh}/K_{mv}=100000)$ & 0.796 & 0.791 & 0.801 \\
$Ri_{\mathrm{new}}$ $(K_{mh}/K_{mv}=1000000)$ & 0.775 & 0.770 & 0.780 \\
\hline
\label{tab:tab4}
\end{tabular}
\end{table}

The second sensitivity study addresses the uncertainty inherent in turbulence reports and how it may affect the evaluation of diagnostic performance. Following the approach of \cite{sharman2006}, the aim is to assess whether the differences in skill between the diagnostics remain robust under resampling of the observations. To quantify this uncertainty, 10\,000\,000 reports were randomly selected from the full MADIS dataset, and 1000 bootstrap replications were performed to estimate the resampling distribution of the AUC for TI1, $Ri_{\mathrm{old}}$, and $Ri_{\mathrm{new}}$ across several values of $K_{mh}/K_{mv}$ (Table~\ref{tab:tab4}). The resulting ensemble of ROC curves forms an “uncertainty envelope” around the mean (Fig.~\ref{fig:fig6}). The uncertainty bands are very tight (approximately $\pm 0.01$ in AUC), but become slightly wider (about $\pm 0.03$) when a smaller number of reports is used (1\,000\,000; Fig.~\ref{fig:fig6A} and Table~\ref{tab:tabA3} in the Appendix).

Across all bootstrap realizations, $Ri_{\mathrm{new}}$ with $K_{mh}/K_{mv}=5000$ consistently yields the highest AUC (mean = 0.848; 95\% CI = 0.843--0.852), TI1 performs slightly lower (AUC = 0.833), and $Ri_{\mathrm{old}}$ exhibits the lowest skill (AUC = 0.819). The confidence intervals for $Ri_{\mathrm{new}}$ at well-performing ratios ($K_{mh}/K_{mv}\in[10^{3},10^{4}]$) do not overlap with those of $Ri_{\mathrm{old}}$: the lower bound of the $Ri_{\mathrm{new}}$ AUC distribution exceeds the upper bounds of both the $Ri_{\mathrm{old}}$ distribution and the TI1 index distribution. This confirms that the improvement provided by $Ri_{\mathrm{new}}$ is statistically robust.

\clearpage
\newpage
\subsection{Seasonal variability}

\begin{figure}[h]
    \centering
    \includegraphics[width=\textwidth]{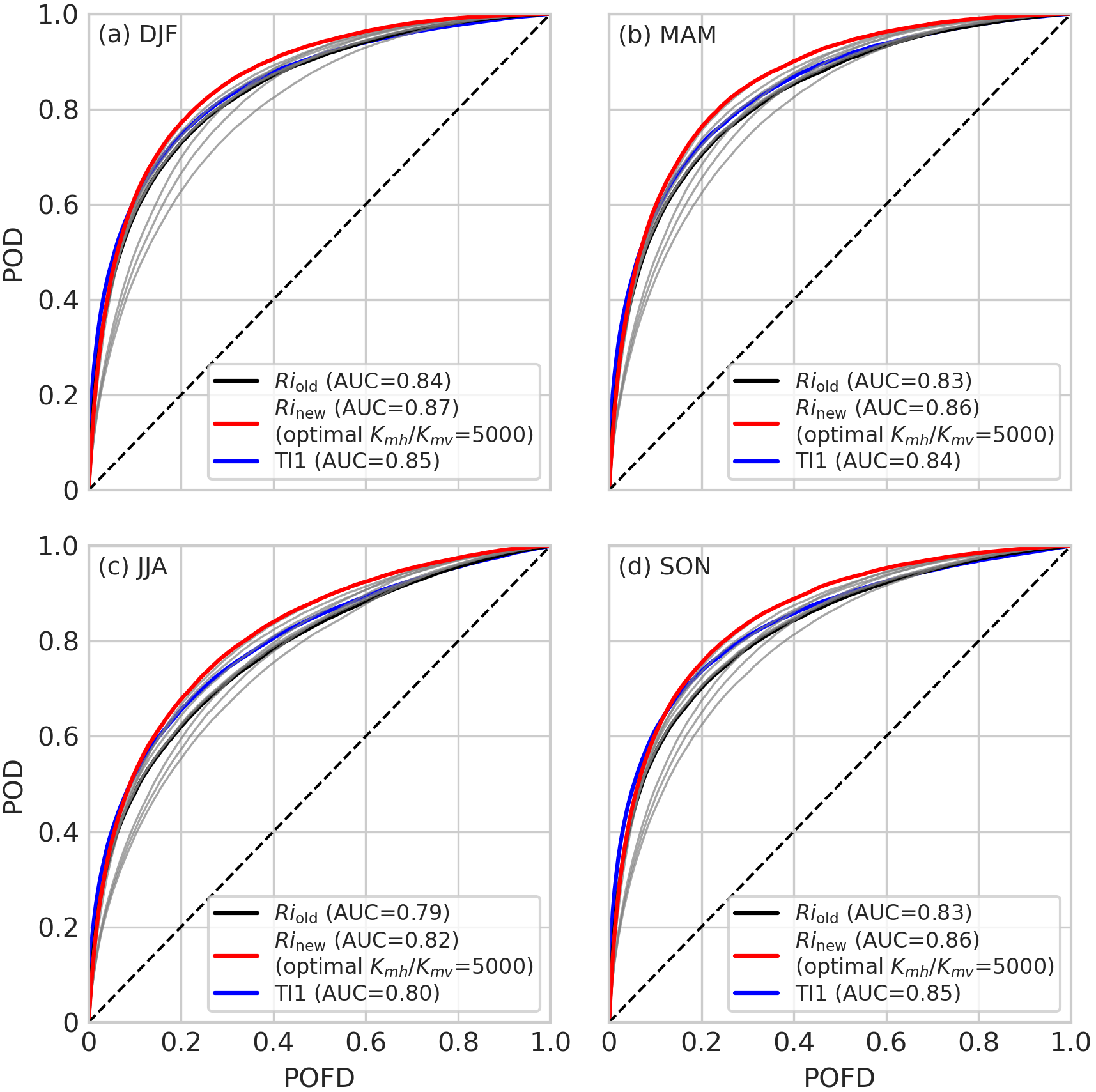}
    \caption{As in Figure~\ref{fig:fig2}, but for (a) winter DJF; (b) spring MAM; (c) summer JJA; (d) autumn SON, for MOG turbulence (EDR~$\ge$~0.20\,m$^{2/3}$\,s$^{-1}$)}
    \label{fig:fig7}
\end{figure}

Seasonal verification for MOG turbulence using an EDR threshold of 0.20~$\mathrm{m}^{2/3} \mathrm{s}^{-1}$ (Fig.~\ref{fig:fig7}) reveals a clear annual cycle in the diagnostic skill. The highest AUC values occur during Northern Hemisphere winter (December--January--February, DJF), when wind shear is more frequent due to the enhanced jet stream and storm-track activity, conditions typically associated with increased clear-air turbulence occurrence. In contrast, the lowest skill is found in summer (June--July--August, JJA), with AUC values reduced by approximately 0.04 for $Ri_{\mathrm{new}}$, 0.06 for $Ri_{\mathrm{old}}$, and 0.05 for TI1 relative to winter. Spring (March--April--May, MAM) and autumn (September--October--November, SON) exhibit intermediate and nearly identical skill, remaining closer to winter than to summer. Sensitivity analyses using different EDR thresholds indicate that the diagnostic skill during winter remains similar, whereas larger variations are observed in summer as the EDR threshold increases (Figs.~\ref{fig:fig7A} and \ref{fig:fig8A}). Despite these seasonal variations in atmospheric conditions, the relative ranking of the diagnostics remains unchanged: $Ri_{\mathrm{new}}$ consistently provides the highest diagnostic skill, followed by TI1 and then $Ri_{\mathrm{old}}$.

In all seasons, $K_{mh}/K_{mv}=5000$ yields the highest AUC and TSS values and remains the best-performing choice (Table~\ref{tab:tab5}). 
The magnitude of improvement relative to $Ri_{\mathrm{old}}$ varies slightly with season, with the largest increase in TSS occurring in summer, when the overall diagnostic skill is weakest, and smaller increases in winter, spring, and autumn. This seasonal dependence is physically consistent. In winter, strong vertical wind shear associated with the jet stream is likely the dominant driver of upper-level turbulence, so the added horizontal-shear terms in $Ri_{\mathrm{new}}$ may provide a more modest benefit. In summer, gravity waves associated with deep convection and other sources may contribute to turbulence through the generation of large spatial gradients, potentially increasing the relevance of the horizontal-shear component. Consequently, the improvement provided by $Ri_{\mathrm{new}}$ is more pronounced in summer than in other seasons. Seasonal variations in the optimal index thresholds are also evident: the TI1 threshold increases from summer to winter, while those for $Ri_{\mathrm{old}}$ and $Ri_{\mathrm{new}}$ decrease, with larger variability observed for TI1 index (Table~\ref{tab:tab5}).

Overall, this seasonal analysis shows that (i) the skill of all diagnostics decreases from winter to summer, consistent with known seasonal variations in turbulence environments, and (ii) the improvement achieved by $Ri_{\mathrm{new}}$ is robust throughout the year. Across all seasons, $Ri_{\mathrm{new}}$ outperforms both $Ri_{\mathrm{old}}$ and TI1 index.

\begin{table}[h]
\centering
\caption{As in Table~\ref{tab:tab2}, but presented separately for each season: winter (DJF), spring (MAM), summer (JJA), and autumn (SON).}
\small

\textbf{DJF}\\[2mm]
\begin{tabular}{lccccc}
\hline\hline
Diagnostic & AUC & POD & PODN & TSS & Threshold \\
\hline
TI1 & 0.849 & 0.743 & 0.804 & 0.548 & $6.04\times10^{-7}$ \\
$Ri_{\mathrm{old}}$ & 0.843 & 0.734 & 0.799 & 0.533 & 1.847 \\
$Ri_{\mathrm{new}}$ $(K_{mh}/K_{mv}=50)$   & 0.845 & 0.732 & 0.802 & 0.534 & 1.814 \\
$Ri_{\mathrm{new}}$ $(K_{mh}/K_{mv}=100)$  & 0.846 & 0.737 & 0.798 & 0.536 & 1.833 \\
$Ri_{\mathrm{new}}$ $(K_{mh}/K_{mv}=500)$  & 0.853 & 0.759 & 0.788 & 0.547 & 1.839 \\
$Ri_{\mathrm{new}}$ $(K_{mh}/K_{mv}=1000)$ & 0.858 & 0.769 & 0.787 & 0.556 & 1.753 \\
$Ri_{\mathrm{new}}$ $(K_{mh}/K_{mv}=5000)$ & \textbf{0.865} & \textbf{0.785} & \textbf{0.791} & \textbf{0.576} & \textbf{1.256} \\
$Ri_{\mathrm{new}}$ $(K_{mh}/K_{mv}=10000)$   & 0.861 & 0.799 & 0.771 & 0.570 & 1.026 \\
\hline
\end{tabular}

\vspace{0.4cm}

\textbf{MAM}\\[2mm]
\begin{tabular}{lccccc}
\hline\hline
Diagnostic & AUC & POD & PODN & TSS & Threshold \\
\hline
TI1 & 0.843 & 0.746 & 0.786 & 0.533 & $5.35\times10^{-7}$ \\
$Ri_{\mathrm{old}}$ & 0.832 & 0.730 & 0.784 & 0.514 & 2.154 \\
$Ri_{\mathrm{new}}$ $(K_{mh}/K_{mv}=50)$  & 0.835 & 0.733 & 0.784 & 0.517 & 2.139 \\
$Ri_{\mathrm{new}}$ $(K_{mh}/K_{mv}=100)$ & 0.837 & 0.735 & 0.785 & 0.519 & 2.124 \\
$Ri_{\mathrm{new}}$ $(K_{mh}/K_{mv}=500)$ & 0.846 & 0.750 & 0.783 & 0.533 & 2.034 \\
$Ri_{\mathrm{new}}$ $(K_{mh}/K_{mv}=1000)$& 0.852 & 0.760 & 0.786 & 0.546 & 1.895 \\
$Ri_{\mathrm{new}}$ $(K_{mh}/K_{mv}=5000)$& \textbf{0.862} & \textbf{0.796} & \textbf{0.774} & \textbf{0.570} & \textbf{1.363} \\
$Ri_{\mathrm{new}}$ $(K_{mh}/K_{mv}=10000)$   & 0.857 & 0.793 & 0.768 & 0.562 & 1.020 \\
\hline
\end{tabular}

\vspace{0.4cm}

\textbf{JJA}\\[2mm]
\begin{tabular}{lccccc}
\hline\hline
Diagnostic & AUC & POD & PODN & TSS & Threshold \\
\hline
TI1 & 0.802 & 0.691 & 0.775 & 0.467 & $4.14\times10^{-7}$ \\
$Ri_{\mathrm{old}}$ & 0.785 & 0.668 & 0.767 & 0.436 & 2.771 \\
$Ri_{\mathrm{new}}$ $(K_{mh}/K_{mv}=50)$  & 0.789 & 0.671 & 0.768 & 0.439 & 2.744 \\
$Ri_{\mathrm{new}}$ $(K_{mh}/K_{mv}=500)$ & 0.804 & 0.705 & 0.755 & 0.460 & 2.660 \\
$Ri_{\mathrm{new}}$ $(K_{mh}/K_{mv}=1000)$& 0.812 & 0.698 & 0.777 & 0.475 & 2.263 \\
$Ri_{\mathrm{new}}$ $(K_{mh}/K_{mv}=5000)$& \textbf{0.822} & \textbf{0.732} & \textbf{0.762} & \textbf{0.495} & \textbf{1.461} \\
$Ri_{\mathrm{new}}$ $(K_{mh}/K_{mv}=10000)$   & 0.815 & 0.726 & 0.757 & 0.483 & 1.019 \\
\hline
\end{tabular}

\vspace{0.4cm}

\textbf{SON}\\[2mm]
\begin{tabular}{lccccc}
\hline\hline
Diagnostic & AUC & POD & PODN & TSS & Threshold \\
\hline
TI1 & 0.845 & 0.718 & 0.830 & 0.548 & $5.78\times10^{-7}$ \\
$Ri_{\mathrm{old}}$ & 0.830 & 0.718 & 0.795 & 0.514 & 2.229 \\
$Ri_{\mathrm{new}}$ $(K_{mh}/K_{mv}=50)$  & 0.833 & 0.723 & 0.793 & 0.517 & 2.237 \\
$Ri_{\mathrm{new}}$ $(K_{mh}/K_{mv}=500)$ & 0.845 & 0.752 & 0.782 & 0.534 & 2.201 \\
$Ri_{\mathrm{new}}$ $(K_{mh}/K_{mv}=1000)$& 0.851 & 0.761 & 0.786 & 0.547 & 2.028 \\
$Ri_{\mathrm{new}}$ $(K_{mh}/K_{mv}=5000)$& \textbf{0.859} & \textbf{0.776} & 0.788 & \textbf{0.564} & \textbf{1.341} \\
$Ri_{\mathrm{new}}$ $(K_{mh}/K_{mv}=10000)$   & 0.853 & 0.782 & 0.772 & 0.554 & 1.018 \\
\hline
\label{tab:tab5}
\end{tabular}
\end{table}

\clearpage
\newpage
\subsection{Regional variability}

\begin{figure}[h]
    \centering
    \includegraphics[width=\textwidth]{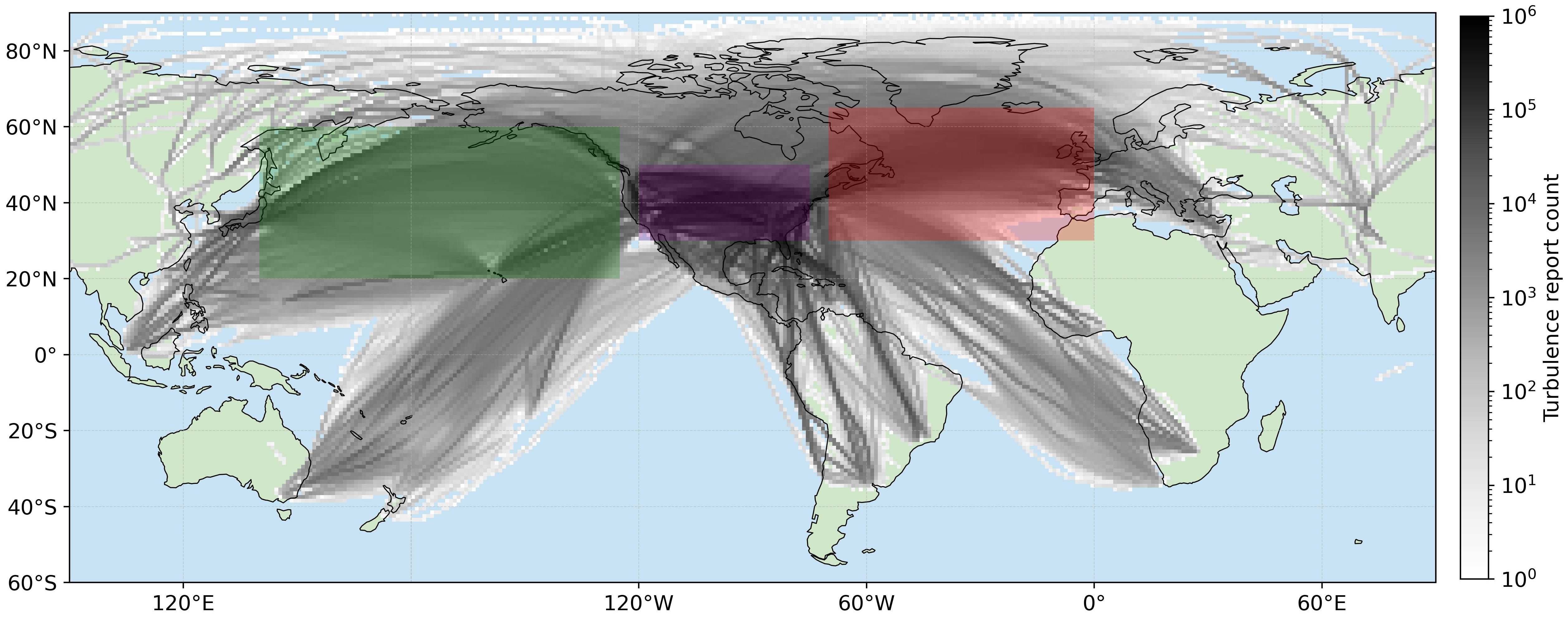}
\caption{
Regions used for the statistical evaluation of the turbulence diagnostics. 
The three analyzed regions are defined as follows: 
(1) \textbf{CONUS} (30--50$^\circ$N, 120--75$^\circ$W; \emph{purple}), 
(2) \textbf{North Atlantic} (30--65$^\circ$N, 70--0$^\circ$W; \emph{red}), 
(3) \textbf{North Pacific} (20--60$^\circ$N, 140$^\circ$E--125$^\circ$W; \emph{green}). 
For the 2017--2024 period, the number of ACARS turbulence reports from the NOAA MADIS archive above 8~km flight altitude within each region is:
CONUS: 120\,752\,371 (48.7\%), 
North Atlantic: 41\,663\,600 (16.8\%), 
and North Pacific: 27\,791\,310 (11.2\%),  
relative to the total number of 247\,730\,014 reports.
}
    \label{fig:fig8}
\end{figure}

Figure~\ref{fig:fig8} shows the three regions used for the regional verification—CONUS (the contiguous United States), the North Atlantic, and the North Pacific—that together represent more than 75\% of all available turbulence reports in the NOAA MADIS archive. About 50\% of these reports are associated with flights over CONUS, meaning that the global performance metrics presented in the previous sections are strongly influenced by atmospheric conditions over CONUS.
It is therefore important to evaluate the diagnostics regionally to assess whether diagnostic skill and ROC characteristics vary across different flow environments.

\begin{figure}[h]
    \centering
    \includegraphics[width=0.35\textwidth]{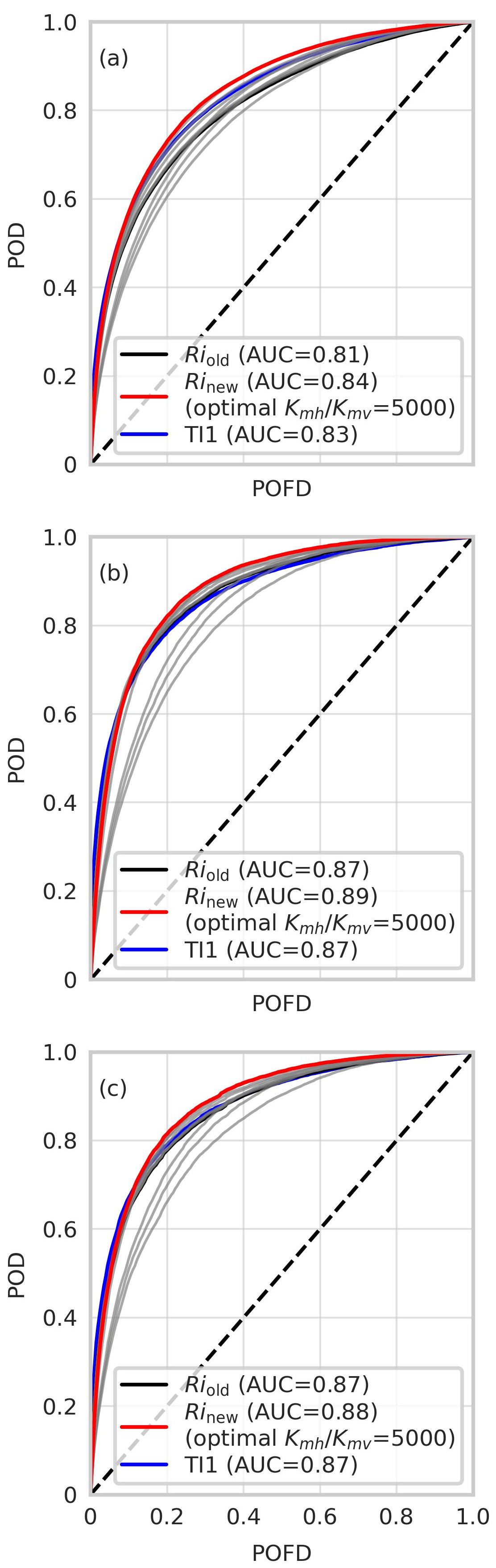}
    \caption{As in Figure~\ref{fig:fig2}, but for (a) CONUS; (b) North Atlantic; (c) North Pacific, for MOG turbulence (EDR~$\ge$~0.20\,m$^{2/3}$\,s$^{-1}$)}
    
    \label{fig:fig9}
\end{figure}

The regional ROC curves for MOG turbulence (EDR $\geq 0.20$~m$^{2/3}$,s$^{-1}$; Fig.~\ref{fig:fig9}) indicate that the relative performance of the TI1 index is higher than that of $Ri_{\mathrm{old}}$ over CONUS, whereas their performance is similar over the oceans (North Atlantic and North Pacific). Again, $Ri_{\mathrm{new}}$ outperforms both TI1 and $Ri_{\mathrm{old}}$ in all regions as measured by the AUC metric. AUC values are highest over the North Atlantic and North Pacific (Figs.~\ref{fig:fig9}b,c) for all diagnostics, consistent with the dominant role of vertical wind shear in turbulence generation over oceans. In these regions, the increase in AUC value for $Ri_{\mathrm{new}}$ relative to $Ri_{\mathrm{old}}$ is approximately 0.02, and about 0.01–0.02 relative to TI1. Over CONUS (Fig.~\ref{fig:fig9}a), AUC values are about 0.04–0.06 lower than over the oceans, while the increase for $Ri_{\mathrm{new}}$ relative to $Ri_{\mathrm{old}}$ is approximately 0.03–0.04, indicating the relevance of including horizontal wind shear in environments where turbulence is generated by a mix of mechanisms rather than purely jet-related vertical shear.

\begin{figure}[h]
    \centering
    \includegraphics[width=\textwidth]{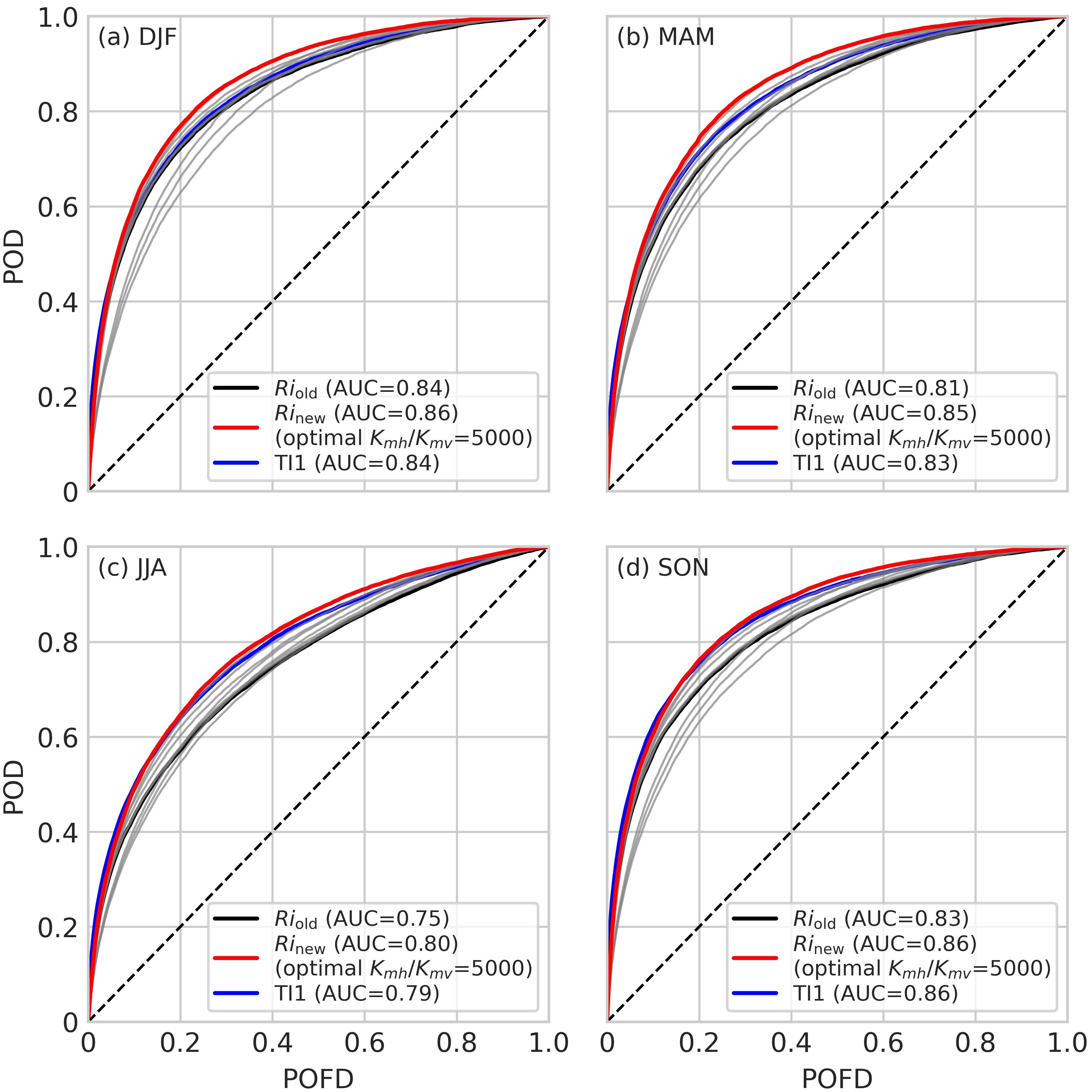}
    \caption{As in Figure~\ref{fig:fig7}, but for CONUS region.}
    \label{fig:fig10}
\end{figure}

\begin{figure}[h]
    \centering
    \includegraphics[width=\textwidth]{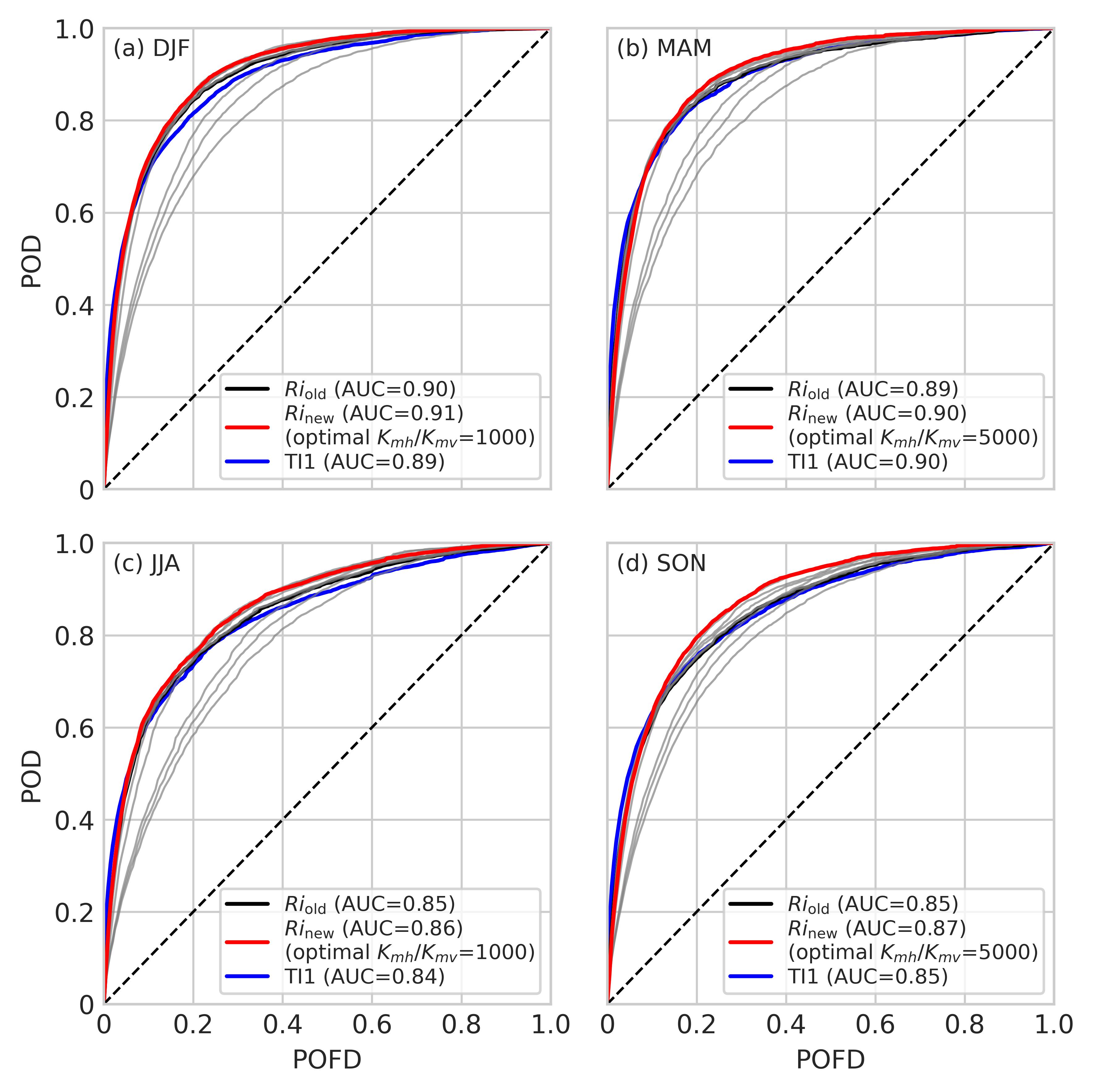}
    \caption{As in Figure~\ref{fig:fig7}, but for the North Atlantic.}
    \label{fig:fig11}
\end{figure}

\begin{figure}[h]
    \centering
    \includegraphics[width=\textwidth]{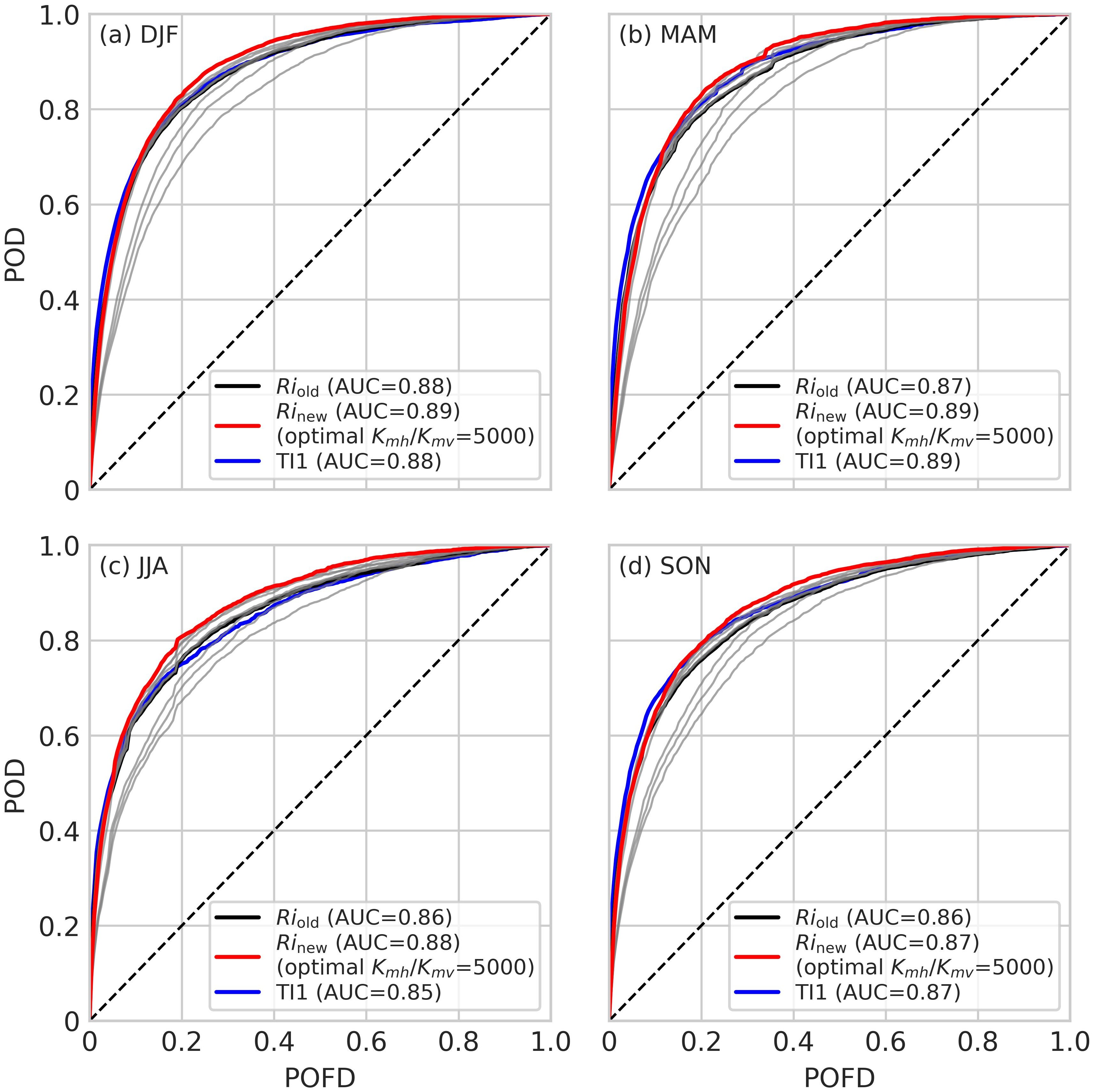}
    \caption{As in Figure~\ref{fig:fig7}, but for the North Pacific.}
    \label{fig:fig12}
\end{figure}

Seasonal verification performed separately for CONUS (Figure~\ref{fig:fig10}), the North Atlantic (Figure~\ref{fig:fig11}), and the North Pacific (Figure~\ref{fig:fig12}) shows again that $Ri_{\mathrm{new}}$ outperforms both TI1 and $Ri_{\mathrm{old}}$ across all seasons and regions. The seasonal cycle in diagnostic skill is strongest over CONUS, where AUC values exhibit a pronounced decrease from winter to summer. For $Ri_{\mathrm{old}}$, the reduction in AUC value between winter and summer is approximately 0.09-0.10, while the decrease is smaller for $Ri_{\mathrm{new}}$ (about 0.05–0.06). The enhanced performance of $Ri_{\mathrm{new}}$ is also most pronounced over CONUS in summer, with AUC increases of approximately 0.05 relative to $Ri_{\mathrm{old}}$ and about 0.01–0.02 relative to TI1 index, consistent with the presence of multiple turbulence-generation mechanisms, including gravity waves generated by deep convection during warm seasons. Over the oceans, the seasonal cycle is much weaker. In the North Pacific, the AUC values of all diagnostics remain high throughout the year, with low variability and $Ri_{\mathrm{new}}$ reaching an AUC of about 0.89. Over the North Atlantic, the seasonality is slightly more pronounced than in the Pacific, consistent with the strong seasonal variability of the jet stream in this region \citep[e.g.,][]{woollings2010}; here, the diagnostics decrease by approximately 0.05 in AUC from winter to summer. In winter, the AUC of $Ri_{\mathrm{new}}$ over the North Atlantic reaches approximately 0.91, and its POD exceeds 80\% for a POFD of 20\%. Overall, regional analysis confirms that while the magnitude of seasonality varies between regions, $Ri_{\mathrm{new}}$ remains the best-performing diagnostic in all regions and seasons.

\clearpage
\newpage
\section{Summary and discussion}
 
A new Richardson number formulation, $Ri_{\mathrm{new}}$, has been developed based on the turbulent kinetic energy budget, explicitly incorporating the contributions of vertical and horizontal wind shear to TKE production. Unlike the classical gradient Richardson number $Ri_{\mathrm{old}}$, which accounts only for vertical wind shear, $Ri_{\mathrm{new}}$ includes horizontal deformation and divergence, weighted by the ratio of horizontal to vertical eddy viscosities ($K_{mh}/K_{mv}$). This new formulation provides a physically based diagnostic for flow regimes in which three-dimensional shear contributes to turbulence production.

The diagnostic skill of $Ri_{\mathrm{new}}$ was evaluated using more than 247 million turbulence reports from 2017--2024. Across a wide range of turbulence intensity thresholds, including light-, moderate-, and severe-or-greater turbulence, $Ri_{\mathrm{new}}$ consistently outperforms both $Ri_{\mathrm{old}}$ and the widely used Turbulence Index~1 (TI1). These improvements are robust across various statistical performance metrics, including ROC-based AUC, probability of detection at operationally relevant false-alarm rates, and the true skill statistic (TSS). On average, $Ri_{\mathrm{new}}$ increases AUC by approximately 0.03–0.05 relative to $Ri_{\mathrm{old}}$, and by about 0.02–0.04 relative to the TI1 index.

Sensitivity studies show that the diagnostic skill of $Ri_{\mathrm{new}}$ is robust to the choice of the ratio of horizontal to vertical eddy viscosities $K_{mh}/K_{mv}$. The highest skill is obtained for values in the range $10^{3}$--$10^{4}$, with peak performance near 5000. This magnitude is consistent with previously reported estimates of $K_{mh}/K_{mv}$ (10$^{4}$–10$^{5}$) in strongly stratified geophysical flows, including the free troposphere and the deep ocean \citep[e.g.,][]{pisso2009,inall2013}. Within this range, diagnostic performance varies only weakly with the parameter value. Additional tests indicate that both diagnostic skill and the value of $K_{mh}/K_{mv}$ that maximizes diagnostic skill are only marginally affected by horizontal grid resolution and by the regridding method, with AUC varying by less than about $\pm 0.01$ across resolutions typical of current numerical weather prediction (NWP) and climate models.

Seasonal and regional analyses reveal differences in diagnostic performance across regions and seasons. Over the oceans, where upper-tropospheric turbulence is mainly associated with strong jet-related vertical wind shear, all diagnostics perform well and exhibit weak seasonal variability. In contrast, over CONUS, where turbulence can be generated by different mechanisms, including 3D shear induced by convection and gravity-wave activity, the added value of $Ri_{\mathrm{new}}$ is more pronounced. The seasonal cycle in diagnostic skill is strongest over CONUS, with AUC values decreasing from winter to summer, particularly for $Ri_{\mathrm{old}}$, by approximately 0.09–0.10. The reduced seasonal sensitivity of $Ri_{\mathrm{new}}$, together with its improved performance, is consistent with the presence of multiple turbulence-generation mechanisms that can produce large horizontal gradients of velocity not fully captured by diagnostics based solely on vertical wind shear, such as $Ri_{\mathrm{old}}$. Although the magnitude of seasonality varies between regions, $Ri_{\mathrm{new}}$ is the best-performing diagnostic in all regions and seasons.

These results highlight the importance of accounting for horizontal wind shear when diagnosing turbulence in the free atmosphere. Although TI1 is a good empirical discriminator of turbulence, its formulation lacks a direct connection to the stabilizing role of stratification. In contrast, $Ri_{\mathrm{new}}$ preserves the physical interpretation of the Richardson number as a balance between stabilizing stratification and shear production, while extending it to explicitly represent three-dimensional shear production.

The value of $K_{mh}/K_{mv}$ that yields the highest diagnostic performance --- approximately 5000 --- provides an indication of the relative weighting of horizontal and vertical shear contributions within the proposed diagnostic at the spatial scales resolved by the ERA5 reanalysis. This magnitude is physically consistent with estimates of $K_{mh}/K_{mv}$ reported for strongly stratified geophysical flows \citep{pisso2009,inall2013}. We note that this value should not be interpreted as a direct estimate of physical eddy viscosities at turbulence scales, but rather as an effective parameter within the closure framework that balances the contributions of vertical and horizontal shear to TKE production at resolved scales.

The relatively large values of this ratio (of order $10^{3}$--$10^{4}$) likely reflect, at least in part, the limited effective resolution of ERA5, which underestimates small-scale velocity 
gradients, particularly in the horizontal. As a result, the contribution of horizontal deformation and divergence to the diagnostic may be underrepresented, requiring a larger weighting factor to capture their role in turbulence generation. This interpretation is consistent with the slight increase of the 
$K_{mh}/K_{mv}$ value that maximizes diagnostic skill when ERA5 is regridded to coarser resolutions (Table~\ref{tab:tab3}), where horizontal averaging further smooths the divergence and deformation terms.

The values reported here are therefore most directly applicable to diagnostics computed from ERA5 or datasets with comparable effective resolution. We note that the World Area Forecast Centres (WAFC London and WAFC Washington) currently issue global aviation turbulence forecasts on a $0.25^\circ$ grid, comparable to the ERA5 horizontal resolution used in this study. At finer resolutions, where horizontal gradients are better resolved, smaller values of this ratio would be expected, and the balance of horizontal and vertical gradients may differ. We therefore caution against applying the ERA5-derived values directly at higher resolutions without re-evaluation.

We acknowledge that the assumption of a spatially uniform $K_{mh}/K_{mv}$ is a simplification. In reality, this ratio is expected to vary with atmospheric stability, flow regime, altitude, and latitude \citep[e.g.,][]{satheesan2002}. A uniform value is adopted here because $K_{mh}/K_{mv}$ is not available from ERA5 or other reanalysis datasets, making a spatially varying formulation impractical at the global scale. The large observational dataset (more than 247 million EDR reports over 
2017--2024, spanning all seasons and regions) shows that the 
identified value is representative of a wide range of atmospheric conditions, and the robustness of the diagnostic skill across seasons and regions (Section~4) further supports this approximation.

Beyond the uniform-value assumption adopted here, a promising direction for future work would be to develop flow-regime-dependent estimates of $K_{mh}/K_{mv}$, analogous to the approach used in the Graphical Turbulence Guidance (GTG) system, in which diagnostic weights are adaptively determined for each forecast cycle \citep{sharman2017}. Such an approach could account for the known dependence of turbulence anisotropy on atmospheric stability and flow regime, and may lead to further improvements in diagnostic skill of $Ri_{\mathrm{new}}$.

It is also important to note that ERA5 does not reproduce the canonical $k^{-5/3}$ mesoscale slope of the kinetic energy spectrum \citep{nastrom1985,lindborg1999}, and its effective resolution is coarser than its nominal $0.25^\circ$ grid \citep{bolgiani2022,li2024}. This is a well-documented limitation for NWP models as well, with effective resolutions typically several times their nominal grid spacing \citep{skamarock2004,abdalla2013,ricard2013,klaver2020}. As a result, the horizontal velocity gradients (divergence and deformation) computed from ERA5 are systematically smaller than the true atmospheric gradients. The large $K_{mh}/K_{mv}$ required by $Ri_{\mathrm{new}}$ is consistent with this resolution-dependent underestimation, and is also physically consistent with the strong anisotropy of stratified geophysical turbulence, where horizontal motions are much more energetic than vertical ones \citep{lilly1983,riley2008, schumann2019}. Applying the diagnostic to a dataset that better resolves the mesoscale spectrum would likely yield a smaller value of this ratio. A detailed evaluation of $Ri_{\mathrm{new}}$ using such datasets is an important direction for future work.

In addition to horizontal resolution, the vertical resolution of the input dataset is also relevant for the computation of turbulence diagnostics, since both static stability ($N^2$) and vertical wind shear depend directly on vertical gradients. ERA5 provides 137 hybrid sigma-pressure levels, with a vertical spacing of approximately 200--300~m in the upper troposphere and lower stratosphere. We note that the vertical gradients relevant here --- vertical wind shear $S_v$ and static stability $N^2$ --- are gradients of the mean flow, not of turbulent fluctuations, 
which are by assumption parameterized. \citet{skamarock2019} showed that the Richardson number probability density function and the kinetic energy spectrum converge in the free atmosphere when the vertical grid spacing is $\Delta z \leq 200$~m, and also noted that the ECMWF IFS (with $\Delta z \approx 300$~m throughout the troposphere, the same configuration as ERA5) likely produces near-converged solutions, in contrast to other operational global models which use coarser vertical mesh spacings of 500--800~m in the same altitude range. The vertical resolution considered here is therefore broadly representative of operational forecast configurations, although some degree of smoothing of vertical gradients is still expected and may influence the absolute magnitude of the diagnostics.

Two aspects are worth noting regarding the influence of vertical resolution on our main conclusions. First, the horizontal-shear terms included in $Ri_{\mathrm{new}}$ (divergence and deformation) do not depend on vertical gradients and are therefore not directly affected by vertical resolution. As a result, vertical resolution primarily influences the vertical-shear and stability terms shared by $Ri_{\mathrm{old}}$ and $Ri_{\mathrm{new}}$, rather than the additional horizontal-shear contribution that distinguishes them. Second, because all diagnostics are computed from the same ERA5 dataset and evaluated against collocated EDR observations, any smoothing associated with vertical resolution affects them consistently, such that the relative performance comparisons presented here are expected to be robust. A full evaluation of $Ri_{\mathrm{new}}$ and other turbulence diagnostics at different vertical resolutions remains to be explored.

A further consideration concerns the reliability of the EDR observations used for verification. EDR is based on the assumption of isotropic and homogeneous turbulence at inertial subrange scales ($\sim$10~m to 1~km), and is derived solely from vertical wind measurements \citep{sharman2014,cornman2016}. Although atmospheric turbulence is often anisotropic in the stably stratified free atmosphere \citep{lilly1983,schumann1995,lane2014}, EDR remains the current standard metric for aviation turbulence verification \citep{ICAO2001,sharman2014,sharman2017}, and all diagnostics evaluated here are verified against the same EDR dataset, ensuring consistency of the relative performance comparisons.

This study compared $Ri_{\mathrm{new}}$ with $Ri_{\mathrm{old}}$ and TI1 to evaluate its added value relative to commonly used turbulence diagnostics. Extending the evaluation to a broader suite of turbulence diagnostics, including those used in the current version of the Graphical Turbulence Guidance (GTG; \cite{sharman2017}), represents an important next step. Because several GTG diagnostics are normalized by $Ri_{\mathrm{old}}$, replacing it with $Ri_{\mathrm{new}}$ within the GTG framework may lead to additional improvements in predictive skill. A comparison of $Ri_{\mathrm{new}}$ with the IFS-CAT index \citep{bechtold2021,ko2025}, once it becomes available as a standard ERA6 output, would also be valuable. Furthermore, the performance of $Ri_{\mathrm{new}}$ for mid- and low-level turbulence, convective regions, as well as mountain-wave turbulence could be investigated; such applications would require retaining the $\overline{w}$ terms in the TKE budget, as the assumption $\overline{w} = 0$ adopted here 
may not hold where vertical motions are substantial.

In summary, this study shows that including horizontal wind shear in the Richardson number yields a physically consistent and statistically robust improvement in diagnosing turbulence in the free atmosphere, relevant to both research and operational aviation applications. The new formulation provides a unified framework for representing the effects of vertical and horizontal wind shear in the stratified atmosphere and for improving turbulence diagnostics in environments where both vertical and horizontal shear are important.

%

%

\clearpage

\acknowledgments
This work was funded by the Leverhulme Trust (Grant RPG-2023-138). The turbulence observations used in this study were made available to the National Oceanic and Atmospheric Administration (NOAA) through contributions from several commercial airlines, including American, Delta, Federal Express, Northwest, United, and United Parcel Service. This research used JASMIN, the UK’s collaborative data analysis infrastructure (\url{https://jasmin.ac.uk}).

%
%
\datastatement
ECMWF’s ERA5 data were accessed and processed from the NERC CEDA archive and are freely available at \url{https://www.ecmwf.int/en/forecasts/datasets/reanalysis-datasets/era5}. ACARS EDR turbulence reports were obtained from the public NOAA MADIS archive and are available at \url{https://madis.ncep.noaa.gov}. The processed data that support the findings of this study are available from the corresponding author upon reasonable request.


%

\clearpage
\newpage

\appendix
\section*{Additional Results and Sensitivity Analyses}

This section provides additional results that support and extend the main analysis, which focuses primarily on MOG turbulence (EDR $\ge 0.20\,\mathrm{m}^{2/3}\,\mathrm{s}^{-1}$). Figures~A2--A6 and Tables~A1--A2 present the results for LOG (EDR $\ge 0.10$) and SOG (EDR $\ge 0.30$) turbulence intensity thresholds, demonstrating that the performance characteristics of $Ri_{\mathrm{new}}$ are broadly consistent across intensity categories. Additional sensitivity tests are also included, such as analyses using other commonly used EDR thresholds (Fig.~A1) and bootstrap resampling experiments to assess statistical robustness (Fig.~A4 and Table~A3).

\begin{figure}[h]
    \centering
    \includegraphics[width=\textwidth]{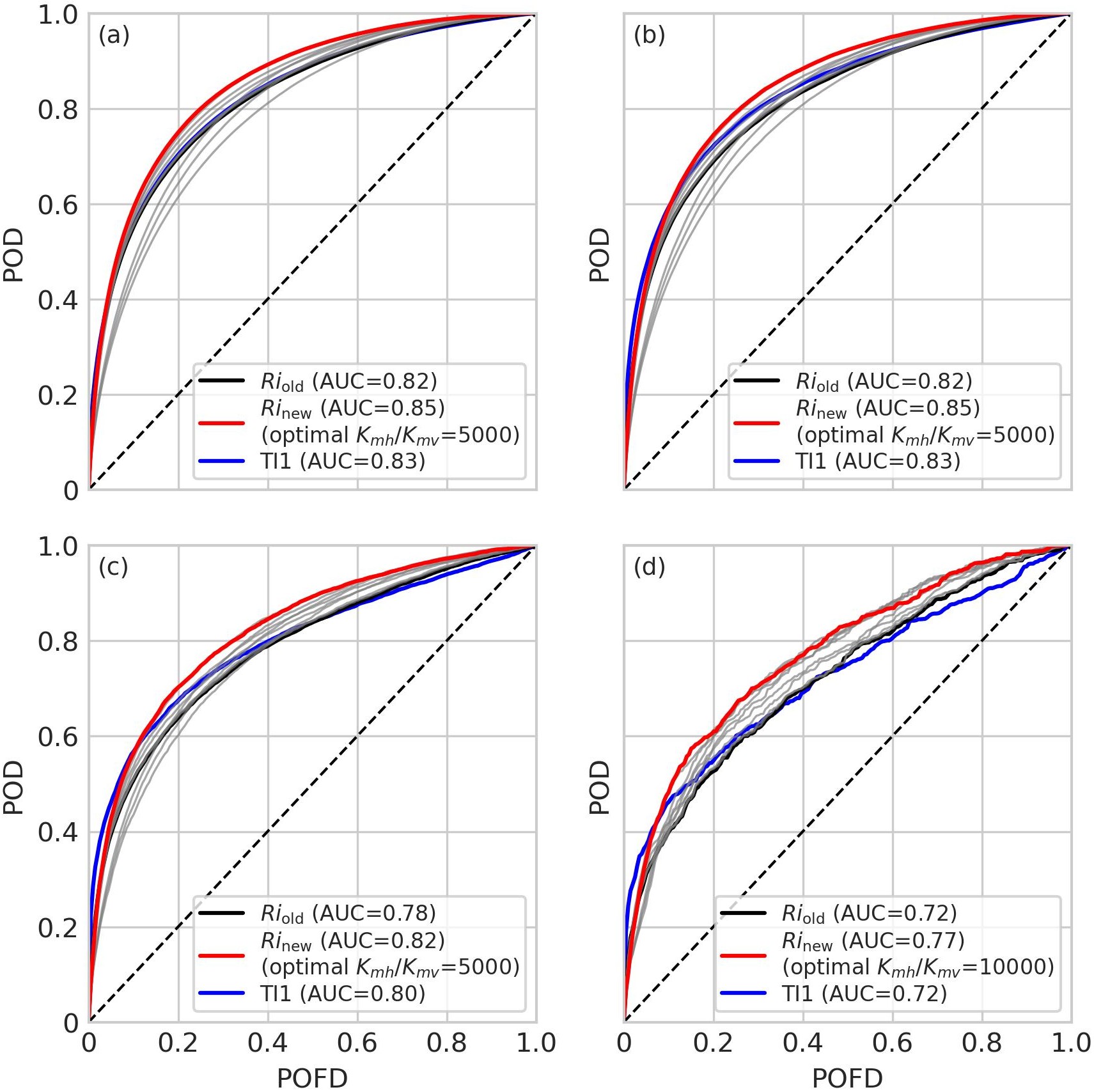}
    \caption{As in Figure~\ref{fig:fig2}, but for (a) EDR $=$ 0.15; (b) EDR $=$ 0.22; (c) EDR $=$ 0.34; (d) EDR $=$ 0.45 $\mathrm{m}^{2/3}\,\mathrm{s}^{-1}$.  The total number of observations used was 247\,730\,014, with 481\,922, 118\,343 , 8\,782, and 930 turbulence events for (a) EDR $\ge$ 0.15, (b) EDR $\ge$ 0.22, (c) EDR $\ge$ 0.34, and (d) EDR $\ge$ 0.45 $\mathrm{m}^{2/3}\,\mathrm{s}^{-1}$, respectively.}
    \label{fig:fig1A}
\end{figure}

\begin{figure}[t]
    \centering
    \includegraphics[width=\textwidth]{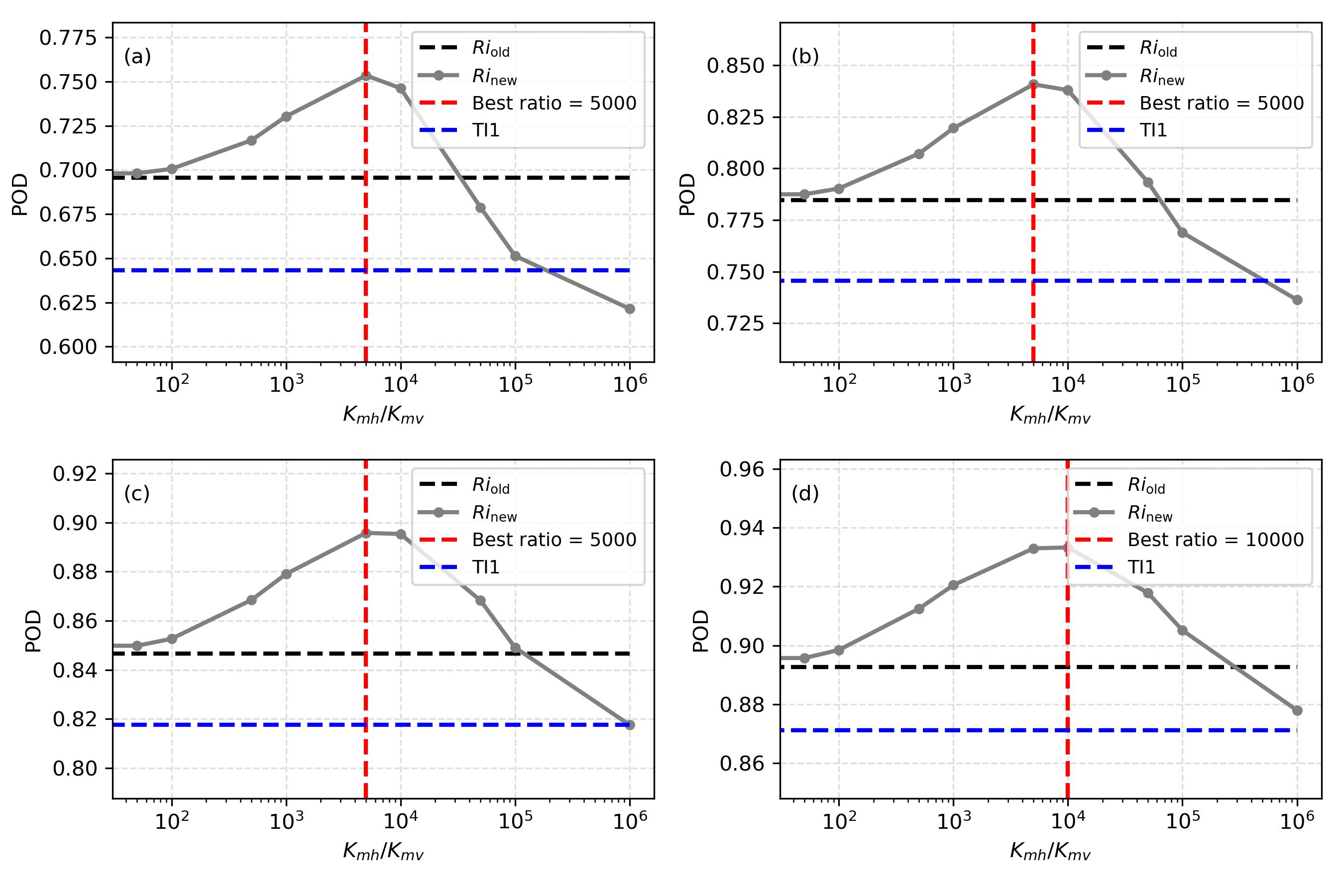}
    \caption{As in Figure~\ref{fig:fig3}, but for LOG turbulence (EDR~$\ge$~0.10\,m$^{2/3}$\,s$^{-1}$)}
    \label{fig:fig2A}
\end{figure}

\begin{figure}[t]
    \centering
    \includegraphics[width=\textwidth]{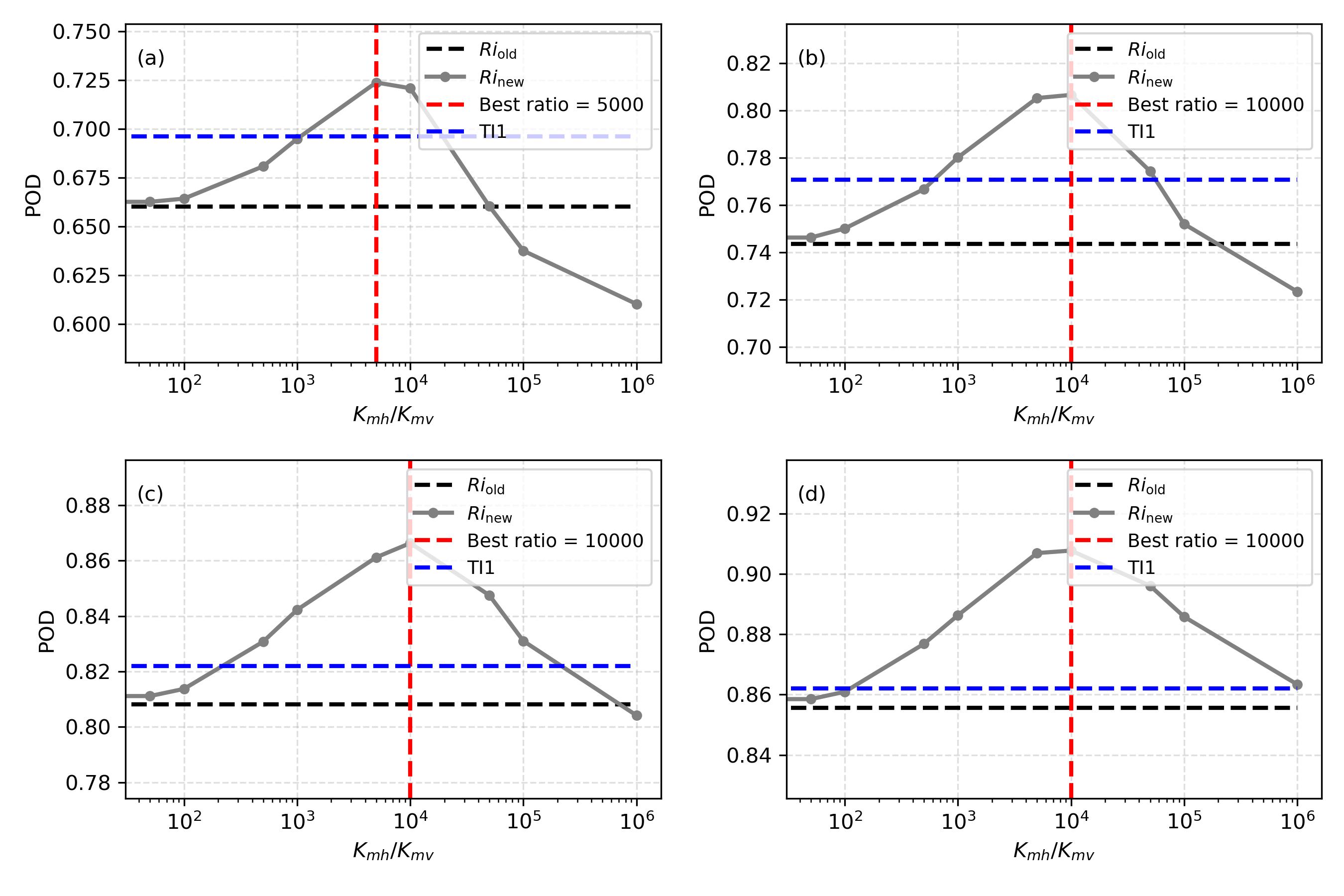}
    \caption{As in Figure~\ref{fig:fig3}, but for SOG turbulence (EDR~$\ge$~0.30\,m$^{2/3}$\,s$^{-1}$)}
    \label{fig:fig3A}
\end{figure}

\begin{figure}[t]
    \centering
    \includegraphics[width=\textwidth]{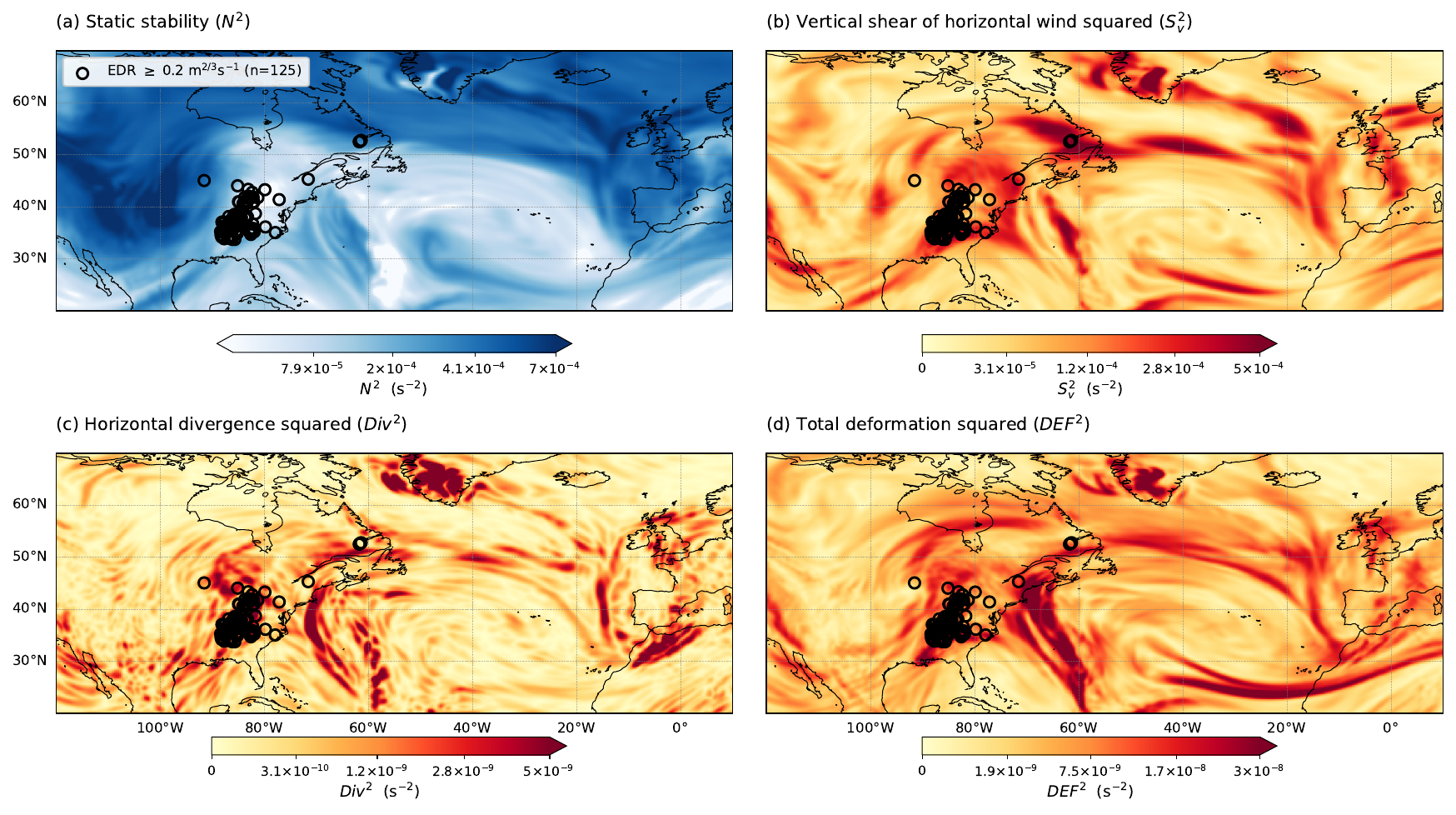}
    \caption{As in Figure~\ref{fig:fig5}, but for 26 March 2024.}
    \label{fig:fig4A}
\end{figure}

\begin{figure}[t]
    \centering
    \includegraphics[width=\textwidth]{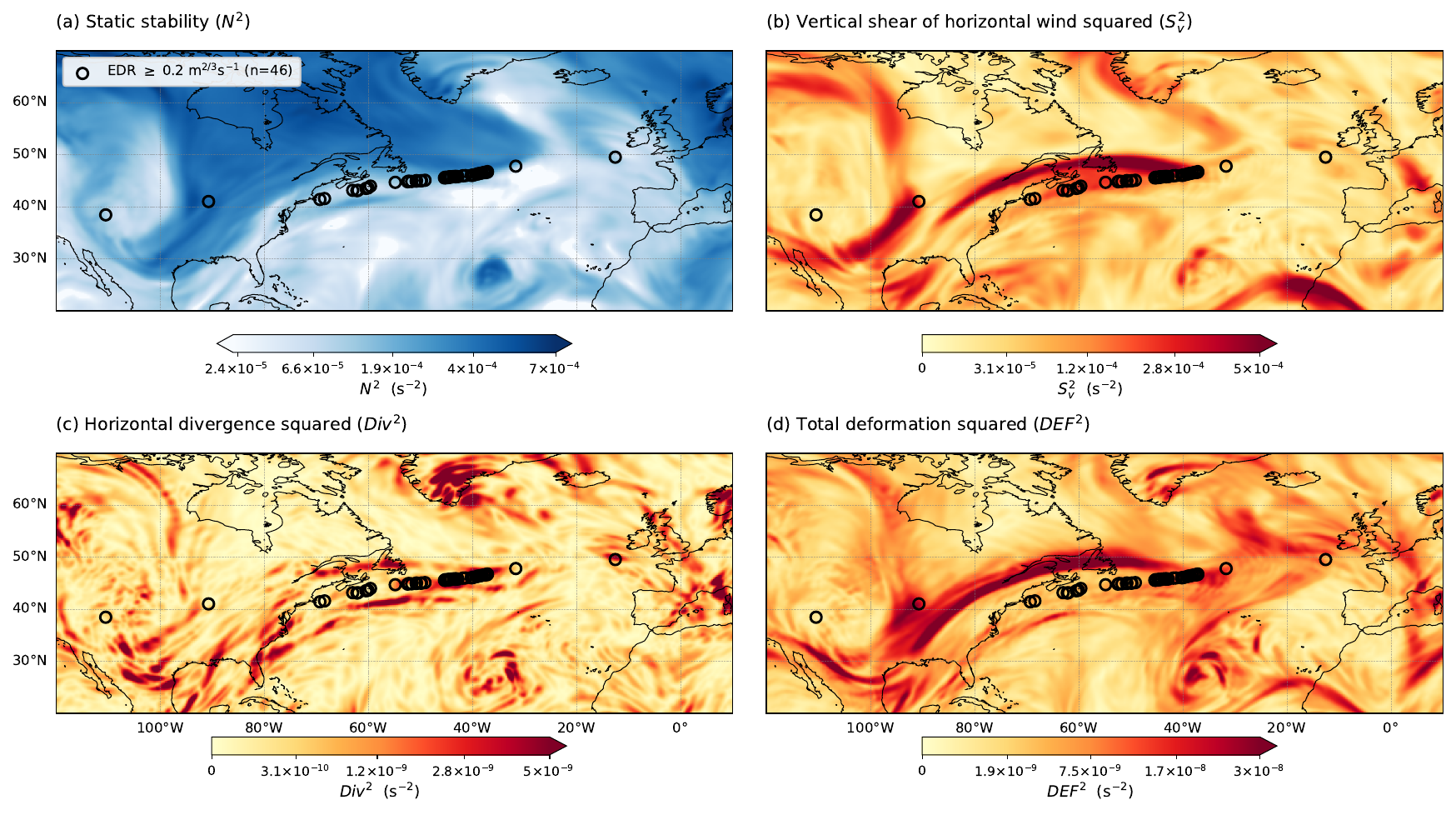}
    \caption{As in Figure~\ref{fig:fig5}, but for 09 December 2017.}
    \label{fig:fig5A}
\end{figure}

\begin{table}[t]
\centering
\caption{As in Table~\ref{tab:tab2}, but for LOG turbulence (EDR~$\ge$~0.10\,m$^{2/3}$\,s$^{-1}$). Metrics shown are the area under the ROC curve (AUC), the probability of detection (POD), the probability of detection of non-events (PODN $=1-\mathrm{POFD}$)), the True Skill Statistic (TSS), and the optimal index threshold (i.e., the value that maximizes TSS). The total number of observations used was 247\,730\,014. The best-performing $Ri_{\text{new}}$ obtained with $K_{mh}/K_{mv}=5000$ is highlighted in bold, as it yields the highest AUC and TSS.}

\small
\begin{tabular}{lccccc}
\hline\hline
Diagnostic & AUC & POD & PODN & TSS & Threshold \\
\hline
$\mathrm{TI1}$                     & 0.799 & 0.711 & 0.744 & 0.455 & $4.33\times10^{-7}$ \\
$Ri_{\mathrm{old}}$               & 0.820 & 0.721 & 0.786 & 0.507 & 2.282 \\
$Ri_{\mathrm{new}}$ $(K_{mh}/K_{mv}=50)$      & 0.832 & 0.725 & 0.785 & 0.509 & 2.273 \\
$Ri_{\mathrm{new}}$ $(K_{mh}/K_{mv}=100)$     & 0.834 & 0.731 & 0.780 & 0.511 & 2.298 \\
$Ri_{\mathrm{new}}$ $(K_{mh}/K_{mv}=500)$     & 0.844 & 0.750 & 0.778 & 0.528 & 2.204 \\
$Ri_{\mathrm{new}}$ $(K_{mh}/K_{mv}=1000)$    & 0.851 & 0.761 & 0.781 & 0.540 & 2.038 \\
$Ri_{\mathrm{new}}$ $(K_{mh}/K_{mv}=5000)$    & \textbf{0.860} & \textbf{0.763} & \textbf{0.782} & \textbf{0.563} & \textbf{1.358} \\
$Ri_{\mathrm{new}}$ $(K_{mh}/K_{mv}=10000)$   & 0.856 & 0.785 & 0.770 & 0.555 & 1.024 \\
$Ri_{\mathrm{new}}$ $(K_{mh}/K_{mv}=50000)$   & 0.825 & 0.757 & 0.717 & 0.495 & 0.399 \\
$Ri_{\mathrm{new}}$ $(K_{mh}/K_{mv}=100000)$  & 0.812 & 0.761 & 0.698 & 0.467 & 0.233 \\
$Ri_{\mathrm{new}}$ $(K_{mh}/K_{mv}=1000000)$ & 0.792 & 0.727 & 0.706 & 0.432 & 0.025 \\
\hline
\label{tab:tabA1}
\end{tabular}
\end{table}

\begin{table}[t]
\centering
\caption{As in Table~\ref{tab:tab2}, but for SOG turbulence (EDR~$\ge$~0.30\,m$^{2/3}$\,s$^{-1}$). Metrics shown are the area under the ROC curve (AUC), the probability of detection (POD), the probability of detection of non-events (PODN $=1-\mathrm{POFD}$), the True Skill Statistic (TSS), and the optimal index threshold (i.e., the value that maximizes TSS). The total number of observations used was 247\,730\,014. The best-performing $Ri_{\text{new}}$ obtained with $K_{mh}/K_{mv}=5000$ is highlighted in bold, as it yields the highest AUC and TSS.}

\small
\begin{tabular}{lccccc}
\hline\hline
Diagnostic & AUC & POD & PODN & TSS & Threshold \\
\hline
$\mathrm{TI1}$                     & 0.826 & 0.676 & 0.847 & 0.523 & $6.36\times10^{-7}$ \\
$Ri_{\mathrm{old}}$               & 0.811 & 0.669 & 0.815 & 0.484 & 1.978 \\
$Ri_{\mathrm{new}}$ $(K_{mh}/K_{mv}=50)$      & 0.813 & 0.672 & 0.816 & 0.486 & 1.964 \\
$Ri_{\mathrm{new}}$ $(K_{mh}/K_{mv}=100)$     & 0.816 & 0.672 & 0.816 & 0.489 & 1.943 \\
$Ri_{\mathrm{new}}$ $(K_{mh}/K_{mv}=500)$     & 0.826 & 0.694 & 0.811 & 0.504 & 1.901 \\
$Ri_{\mathrm{new}}$ $(K_{mh}/K_{mv}=1000)$    & 0.833 & 0.720 & 0.798 & 0.518 & 1.881 \\
$Ri_{\mathrm{new}}$ $(K_{mh}/K_{mv}=5000)$    & \textbf{0.845} & \textbf{0.752} & \textbf{0.789} & \textbf{0.540} & \textbf{1.307} \\
$Ri_{\mathrm{new}}$ $(K_{mh}/K_{mv}=10000)$   & 0.842 & 0.769 & 0.766 & 0.535 & 1.024 \\
$Ri_{\mathrm{new}}$ $(K_{mh}/K_{mv}=50000)$   & 0.815 & 0.765 & 0.713 & 0.479 & 0.399 \\
$Ri_{\mathrm{new}}$ $(K_{mh}/K_{mv}=100000)$  & 0.803 & 0.763 & 0.689 & 0.453 & 0.238 \\
$Ri_{\mathrm{new}}$ $(K_{mh}/K_{mv}=1000000)$ & 0.785 & 0.730 & 0.692 & 0.422 & 0.026 \\
\hline
\label{tab:tabA2}
\end{tabular}
\end{table}

\begin{figure}[t]
    \centering
    \includegraphics[width=0.5\textwidth]{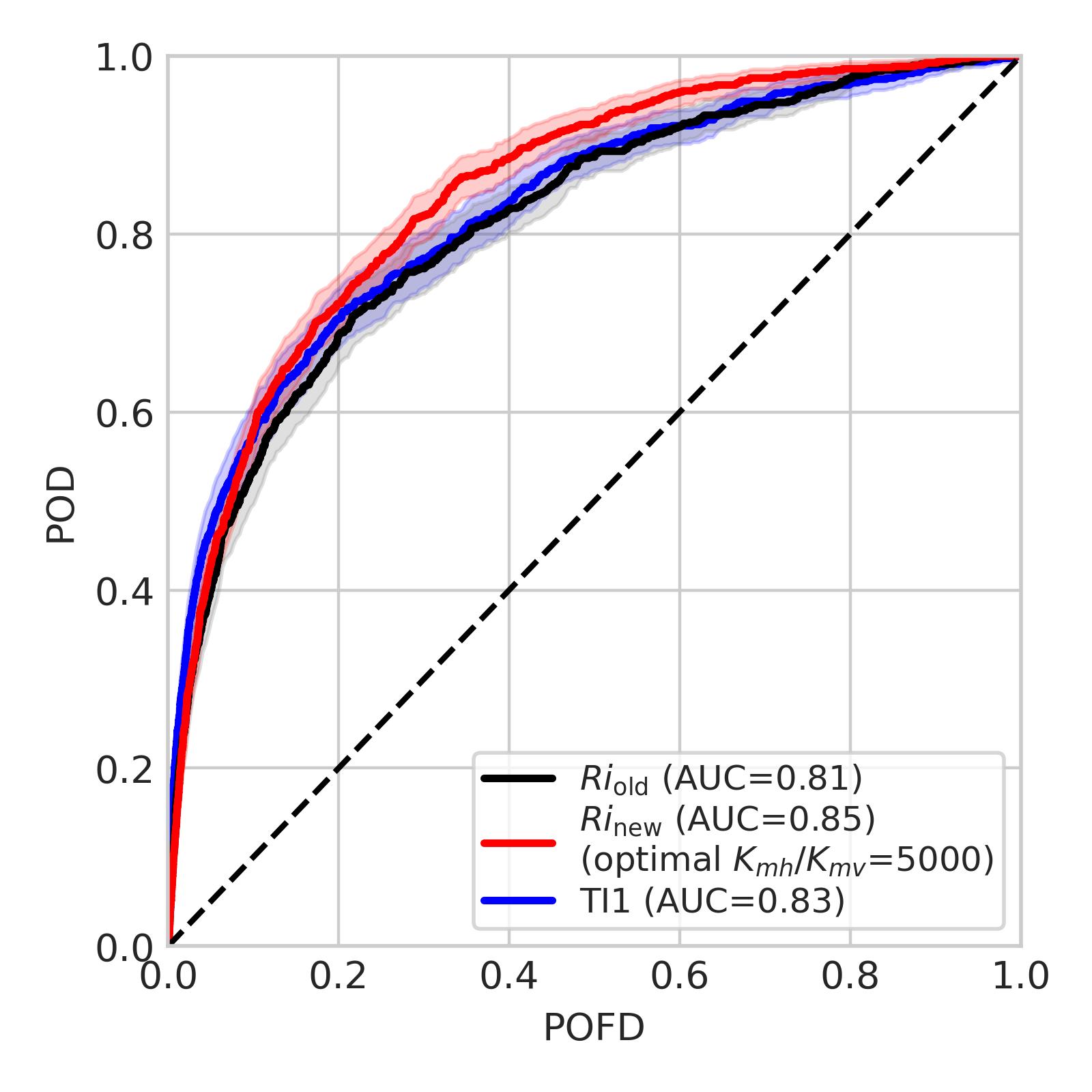}
    \caption{As in Figure~\ref{fig:fig6}, but using a subsample of 1\,000\,000 reports, and  1000 bootstrap replications.}
    \label{fig:fig6A}
\end{figure}

\begin{figure}[t]
    \centering
    \includegraphics[width=\textwidth]{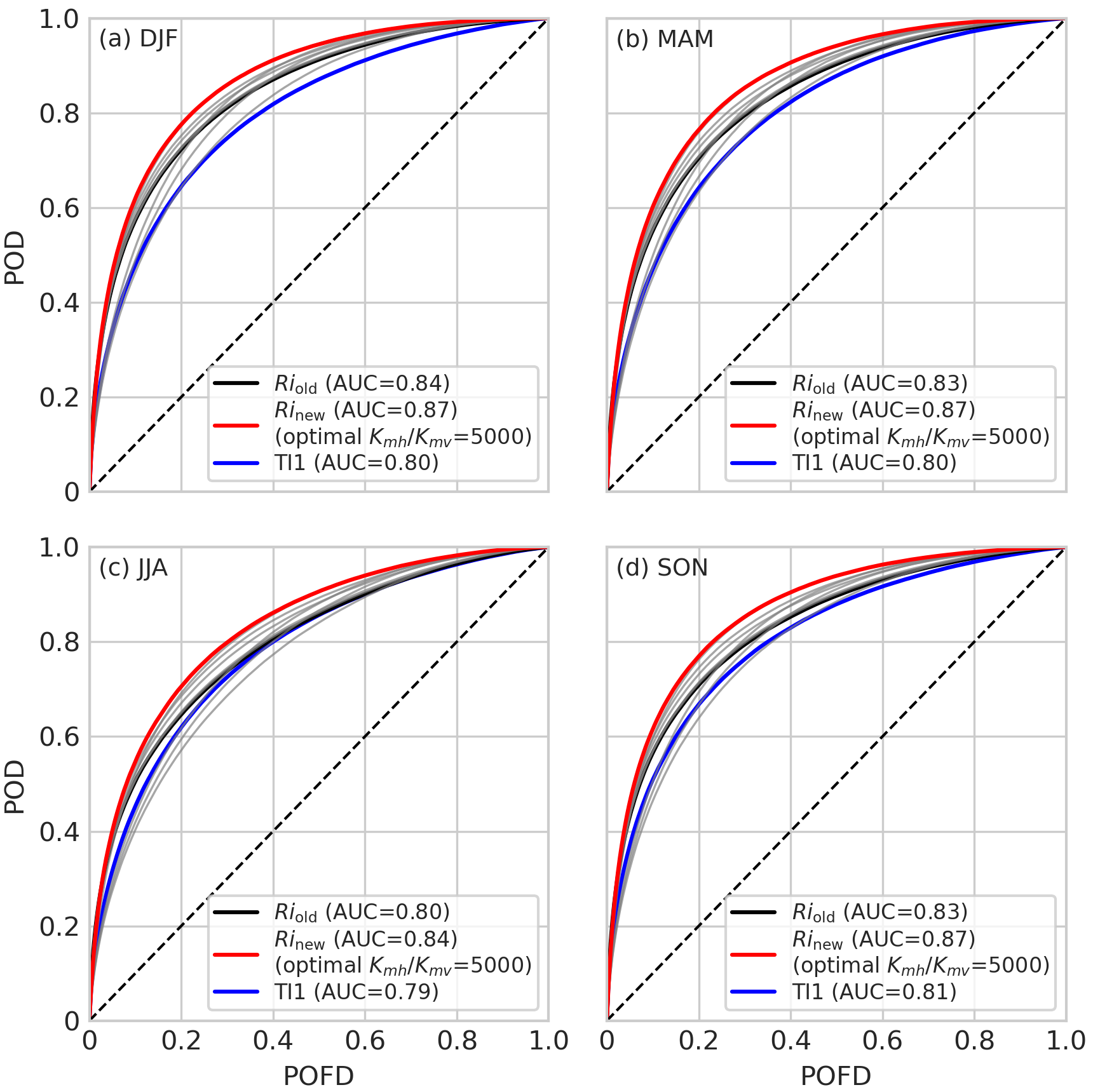}
    \caption{As in Figure~\ref{fig:fig7}, but for LOG turbulence (EDR~$\ge$~0.10\,m$^{2/3}$\,s$^{-1}$)}
    \label{fig:fig7A}
\end{figure}

\begin{figure}[t]
    \centering
    \includegraphics[width=\textwidth]{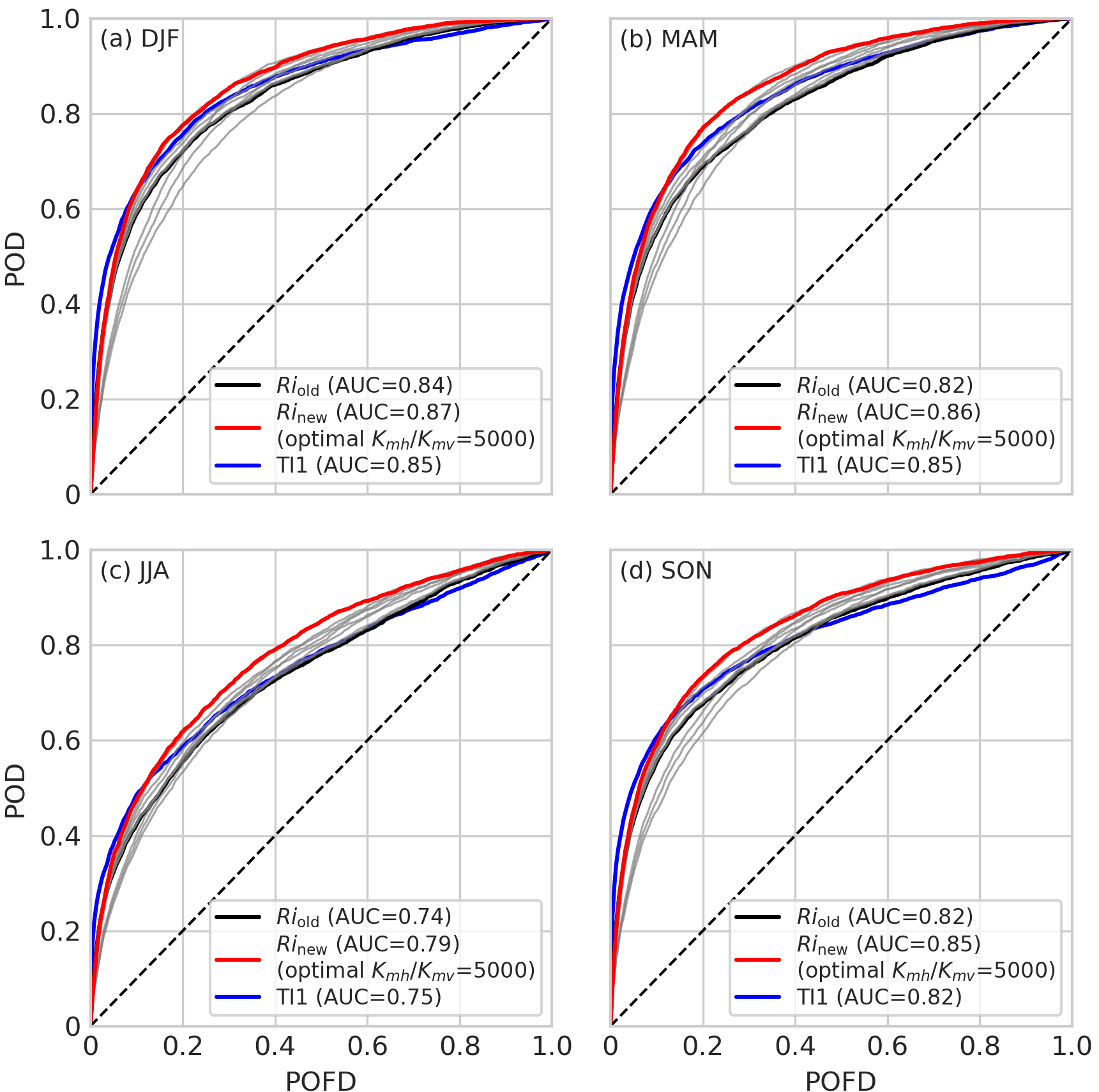}
    \caption{As in Figure~\ref{fig:fig7}, but for SOG turbulence (EDR~$\ge$~0.30\,m$^{2/3}$\,s$^{-1}$)}
    \label{fig:fig8A}
\end{figure}

\begin{table}[t]
\centering
\caption{As in Table~\ref{tab:tab4}, but using a subsample of 1\,000\,000 reports, and  1000 bootstrap replications.}

\small
\begin{tabular}{lccc}
\hline\hline
Diagnostic & AUC (mean) & AUC (2.5\%) & AUC (97.5\%) \\
\hline
TI1 & 0.826 & 0.810 & 0.842 \\
$Ri_{\mathrm{old}}$ & 0.813 & 0.797 & 0.829 \\
$Ri_{\mathrm{new}}$ $(K_{mh}/K_{mv}=50)$ & 0.818 & 0.802 & 0.832 \\
$Ri_{\mathrm{new}}$ $(K_{mh}/K_{mv}=100)$ & 0.820 & 0.803 & 0.835 \\
$Ri_{\mathrm{new}}$ $(K_{mh}/K_{mv}=500)$ & 0.830 & 0.816 & 0.846 \\
$Ri_{\mathrm{new}}$ $(K_{mh}/K_{mv}=1000)$ & 0.837 & 0.821 & 0.852 \\
\textbf{$Ri_{\mathrm{new}}$ $(K_{mh}/K_{mv}=5000)$} 
  & \textbf{0.846} & \textbf{0.832} & \textbf{0.860} \\
$Ri_{\mathrm{new}}$ $(K_{mh}/K_{mv}=10000)$ & 0.842 & 0.828 & 0.855 \\
$Ri_{\mathrm{new}}$ $(K_{mh}/K_{mv}=50000)$ & 0.810 & 0.796 & 0.824 \\
$Ri_{\mathrm{new}}$ $(K_{mh}/K_{mv}=100000)$ & 0.798 & 0.783 & 0.812 \\
$Ri_{\mathrm{new}}$ $(K_{mh}/K_{mv}=1000000)$ & 0.778 & 0.760 & 0.794 \\
\hline
\label{tab:tabA3}
\end{tabular}
\end{table}





%




\clearpage
\newpage
\bibliographystyle{ametsocV6}
\bibliography{references}

\end{document}